\PassOptionsToPackage{table}{xcolor}
\documentclass{article}
\usepackage[T1]{fontenc}
\usepackage{iclr2027_conference,times}
\usepackage{graphicx}
\usepackage{algorithm}
\usepackage{algorithmic}
\usepackage{float}
\usepackage{amsmath,amssymb}
\usepackage{booktabs,multirow,array,arydshln}
\usepackage[most]{tcolorbox}
\usepackage{placeins}
\usepackage{capt-of}
\usepackage{hyperref}
\usepackage{url}
\usepackage{wrapfig}
\usepackage{caption}
\usepackage{subcaption}
\usepackage[sort&compress]{cleveref}
\usepackage{bibunits}

\defaultbibliographystyle{iclr2027_conference}
\defaultbibliography{references}

\makeatletter
\newcommand{\bibunitprefix}{main.}

\renewcommand*{\hyper@natlinkstart}[1]{%
  \Hy@backout{#1}%
  \hyper@linkstart{cite}{cite.\bibunitprefix#1}%
  \def\hyper@nat@current{#1}}

\renewcommand*{\hyper@natlinkbreak}[2]{%
  \hyper@linkend#1%
  \hyper@linkstart{cite}{cite.\bibunitprefix#2}}

\renewcommand*{\hyper@natanchorstart}[1]{%
  \Hy@raisedlink{%
    \hyper@anchorstart{cite.\bibunitprefix#1}}}

\makeatother

\crefname{table}{Table}{Tables}
\crefrangeformat{table}{#3#1#4--#5#2#6}

\newtheorem{definition}{Definition}

\title{Imagine Before Acting: Prospective Skill Identification for LLM Agents}
\author{
    Shuo Liu\textsuperscript{\rm 1}\thanks{Equal contribution.}\hspace{0.45em},
    Yutong Yang\textsuperscript{\rm 1}\footnotemark[1]\hspace{0.45em},
    Haohao Xiao\textsuperscript{\rm 1}\hspace{0.05em},
    Mouxing Yang\textsuperscript{\rm 1}\hspace{0.05em},
    Xi Peng\textsuperscript{\rm 2,3} \\
    \textsuperscript{\rm 1}College of Computer Science, Sichuan University \\
    \textsuperscript{\rm 2}School of Artificial Intelligence, Sichuan University \\
    \textsuperscript{\rm 3}National Key Laboratory of Fundamental Algorithms and Models for Engineering Numerical \\ Simulation, Sichuan University
}

\iclrfinalcopy 

\begin{document}

\begin{bibunit}

\maketitle
\lhead{}
\addtocontents{toc}{\protect\setcounter{tocdepth}{-1}}

\begin{abstract}
Effective skill identification is crucial for LLM agents to reuse task-solving experience for complex task execution, which however remains challenging as agents must accurately identify useful skills from a large skill gallery and invoke them on demand.
Existing studies have explored various strategies for skill retrieval and invocation, but largely overlook the natural discrepancy between task descriptions and skill documents.
To be specific, our empirical study reveals that task descriptions mainly specify task objectives, whereas skill documents describe the required capabilities and procedural guidance for achieving the objectives.
Such a so-called Task--Skill Misalignment (TSM) problem would cause desirable skills to be neglected during retrieval or remain unselected during invocation, thus hindering task execution.
As a remedy, inspired by human prospective cognition, we propose SkillDreamer, a prospective skill identification framework to mitigate TSM.
In brief, SkillDreamer first infers execution-required capabilities and imagines their realizations as pseudo skills, thereby translating task objectives into execution-oriented guidance for skill retrieval.
As execution proceeds, SkillDreamer reevaluates the capabilities required by emerging subtasks to determine which skills should be invoked, thereby aligning skill invocation with evolving execution needs.
Extensive experiments on three representative agent skill benchmarks not only verify the effectiveness of SkillDreamer in mitigating TSM, but also demonstrate its generalizability across diverse retrievers and LLM agents.
The code will be released upon acceptance.
\end{abstract}

\section{Introduction}
Large language model (LLM) agents have emerged as a promising paradigm for solving complex tasks through autonomous planning~\citep{react}, external tool invocation~\citep{toolformer}, and dynamic interaction with environments~\citep{voyager}.
However, solving each incoming task from scratch requires agents to repeatedly devise execution plans and rely solely on their inherent capabilities, limiting both execution efficiency and reliability.
As a remedy,~\citet{agentskills} proposes encapsulating task-specific workflows, domain expertise, and supporting resources into structured documents, termed \textit{Skills}, whereby agents could access and reuse task-solving knowledge on demand.
As human-authored and automatically generated skills rapidly accumulate in large skill libraries, identifying the most useful skills for given tasks becomes increasingly challenging, making effective skill identification as one of the most fundamental challenges for agents.

Toward effective skill identification, existing agents generally follow a two-stage paradigm, \textit{i.e.}, skill retrieval and skill invocation.
Specifically, given an overall task description, the agent first retrieves a set of candidate skills from the skill gallery to support subsequent execution.
As execution proceeds, the agent progressively plans and decomposes the overall task into concrete subtasks, invoking relevant skills from the candidate set on demand.
Within this paradigm, prior studies have explored different strategies for skill retrieval and invocation.
On the one hand, during skill retrieval, existing methods focus on improving retrieval performance by training skill-customized retrievers~\citep{skillret}, adopting multi-stage retrieve-and-rerank pipelines~\citep{skillrouter}, or decomposing complex tasks for finer-grained matching~\citep{SKillRerank,compskill}. 
On the other hand, during skill invocation, existing methods aim to improve the accuracy of selecting desirable skills by resorting to the current environment state and interaction history~\citep{asi,sgdr,cuaskill}, or learn policies for whether and when to invoke skills~\citep{selskill}.

\begin{figure}[!t]
    \centering
    \includegraphics[width=1\linewidth]{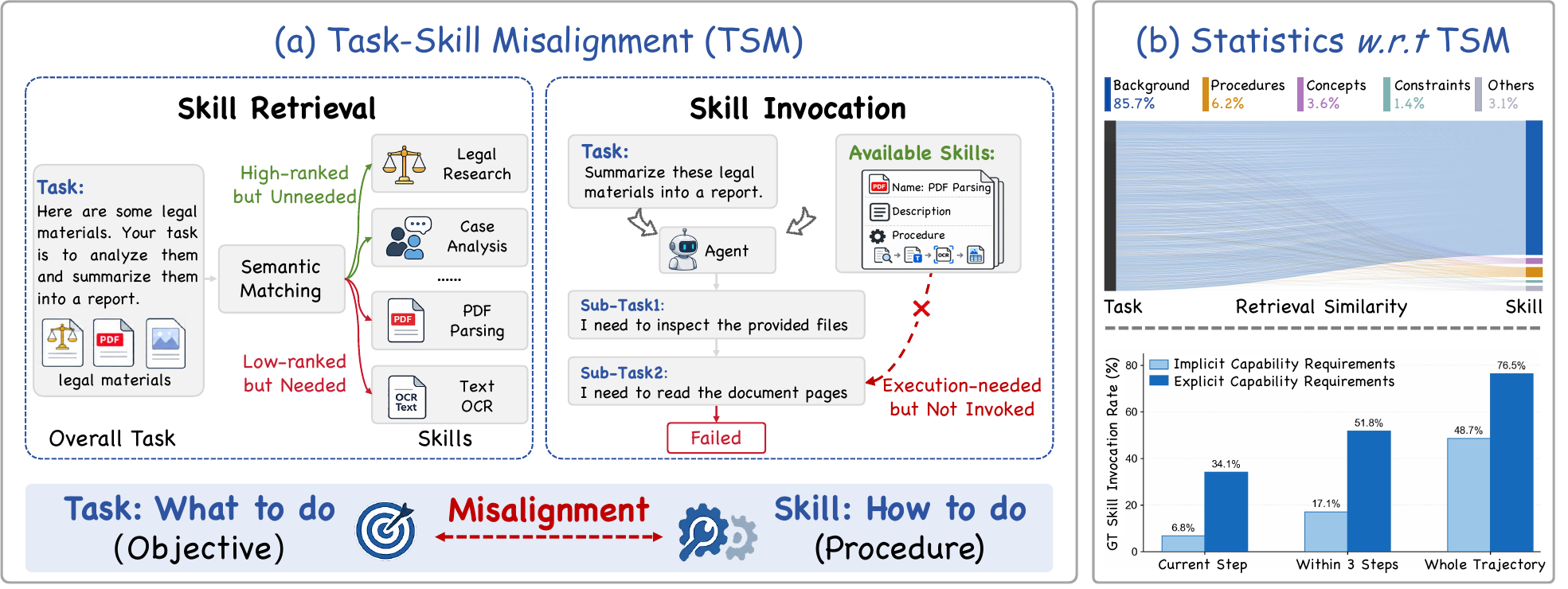}
\caption{
\textbf{(a) Our observation.} 
Task--Skill Misalignment: tasks mainly specify desired objectives, while skills usually manifest as the procedural knowledge required for execution.
As a result, execution-critical skills might be ranked lower during retrieval or remain uninvoked during execution.
\textbf{(b) Statistics \textit{w.r.t} TSM.}
Top: task--skill matching is dominated by background information, while execution-related procedures contribute much less.
Bottom: explicit capability requirements are associated with higher invocation rates of ground-truth skills.
In this figure, we use Qwen3-Embedding-4B to measure the similarity between the query task and different components of skills on the SkillRet benchmark.
For invocation, we provide DeepSeek-V4-Flash with ground-truth skills on SkillUsage and use GPT-6-Sol to evaluate whether capability requirements are explicitly expressed in the execution trajectories.
}
    \label{fig:qsm}
\end{figure}

Despite their promising performance, existing studies always neglect the gap between task descriptions and skill documents, leading to the so-called Task--Skill Misalignment (TSM) problem.
To be specific, as illustrated in Fig.~\ref{fig:qsm}(a), task mainly specify the objectives of users (\textit{e.g.}, generating reports) whereas skills typically describe the capabilities and execution guidance required to achieve the objectives (\textit{e.g.}, PDF Parsing). 
As shown in Fig.~\ref{fig:qsm}(b), the TSM problem could hinder an incentive to significantly hinder the correct identification for desirable skill during both the skill retrieval and invocation stages.
To bridge this gap, we draw inspiration from human \textit{prospective cognition}, which involves imagining future situations in advance to guide current decisions~\citep{prospection}.
Similarly, effective skill identification requires looking beyond task objectives to imagine what execution will demand. 
Therefore, we aim to reason about the capabilities needed to accomplish a task or subtask and render these requirements explicit, bridging the gap between task objectives and execution-oriented skills across both retrieval and invocation.

Based on the above observations and motivations, we propose SkillDreamer, a prospective skill-identification framework for mitigating TSM across skill retrieval and invocation.
In brief, SkillDreamer consists of three modules, \textit{i.e.}, Prospective Skill Generation (PSG), Hybrid Skill Retrieval (HSR), and Capability-aware Skill Invocation (CSI).
Given a task, PSG first infers the required capabilities and imagines how they could be realized, formulating the anticipated procedural guidance as pseudo skills. 
After that, HSR combines these pseudo skills with the original task description to retrieve candidate skills, accounting for both task relevance and execution needs. During execution, CSI continually reevaluates the capabilities required by emerging subtasks and determines whether and which candidate skills to invoke. 

In summary, the main contributions and novelties of this work could be summarized as follows:
\begin{itemize}
    \item We study a practical yet less-explored prblem termed Task--Skill Misalignment (TSM) in agent skill identification, and empirically reveal that the discrepancy between task requirements and execution-oriented skills could significantly hinder both skill retrieval and invocation.
    \item Inspired by prospective cognition of humans, we propose SkillDreamer, a prospective skill-identification framework that anticipates required capabilities and their realization strategies for retrieval, while further performing capability-aware skill invocation during execution.
    \item We propose SkillDreamer, a novel framework which unifies skill retrieval and invocation through prospective reasoning over explicit capability requirements. By translating task objectives into imagined procedural guidance for retrieval and exposing emerging capability needs for invocation, SkillDreamer could bridge task-side intent and skill-side execution knowledge across both stages.
    \item Extensive experiments on three representative skill benchmarks verify the effectiveness of SkillDreamer in both skill retrieval and end-to-end task execution.
    Further evaluations across multiple retrievers and LLM agents establish its broad applicability.
\end{itemize}
\section{Method}

\begin{figure*}[t]
    \centering
    \includegraphics[width=\linewidth]{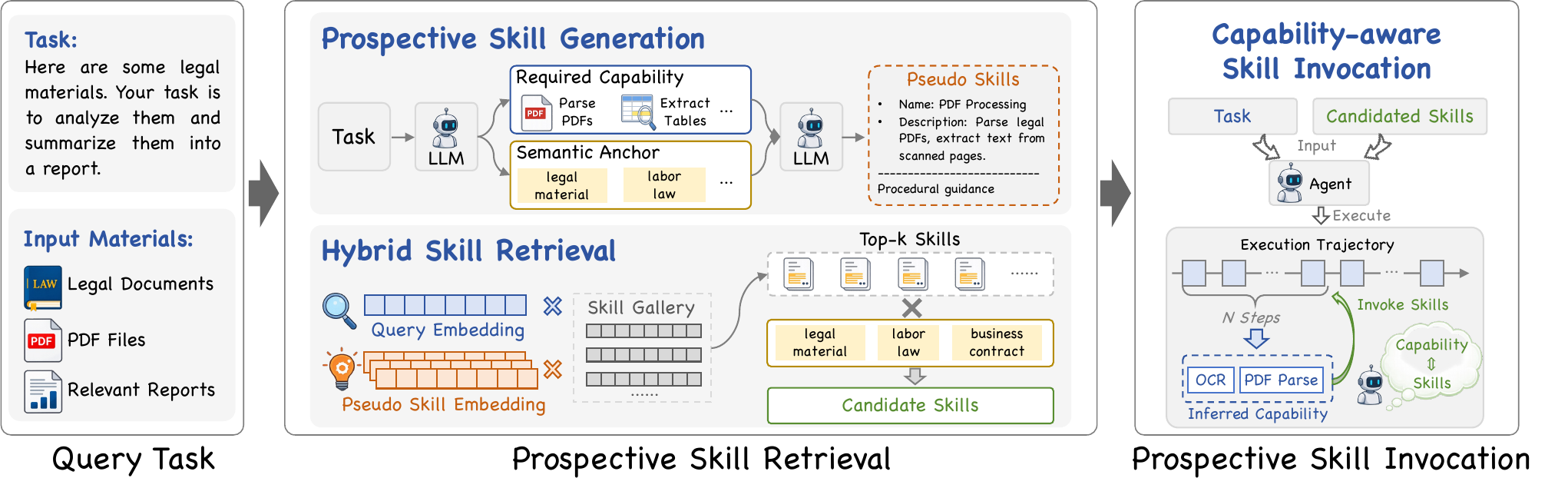}
    \caption{
    \textbf{Overview of the proposed SkillDreamer.}
    Given a task, PSG anticipates the capabilities required for execution and generates corresponding pseudo skills while preserving task-specific semantics.
    HSR then combines the original task with the generated pseudo skills to retrieve a set of candidate skills.
    During execution, CSI reassesses the capability required by the current subtask before action and invokes an appropriate candidate skill when needed.
    }
    \label{fig:overview}
\end{figure*}

In this section, we present SkillDreamer for mitigating Task--Skill Misalignment (TSM) in skill retrieval and invocation.
As illustrated in Fig.~\ref{fig:overview}, SkillDreamer consists of two stages, namely, Prospective Skill Retrieval and Prospective Skill Invocation.
The section is organized as follows.
In Sec.~\ref{sec:problem_formulation}, we first formally define TSM in both stages.
In Sec.~\ref{sec:prospective_generation}, we introduce Prospective Skill Generation (PSG) to infer task-required capabilities and construct corresponding pseudo skills.
In Sec.~\ref{sec:hybrid_retrieval}, we present Hybrid Skill Retrieval (HSR) to retrieve candidate skills by jointly considering the original task and prospective execution information.
Finally, in Sec.~\ref{sec:capability_invocation}, we introduce Capability-aware Skill Invocation (CSI) to identify capability requirements during execution and invoke appropriate skills on demand.

\subsection{Problem Formulation}
\label{sec:problem_formulation}

Given a task $q$ and a skill gallery $\mathcal{S}=\{s_i\}_{i=1}^{G}$, agent skill identification generally involves two stages, \textit{i.e.}, skill retrieval before execution and skill invocation during execution.
At the retrieval stage, a retriever $\mathcal{R}$ ranks the gallery skills according to the overall task and selects the top-$K$ candidates for subsequent execution:
\begin{equation}
    \mathcal{C}
    =
    \mathcal{R}
    \left(
        q,\mathcal{S}
    \right),
    \qquad
    |\mathcal{C}|=K,
\end{equation}
where $\mathcal{C}$ denotes the retrieved candidate skill set.

During task execution, the agent progressively plans and decomposes the overall task into a sequence of concrete subtasks.
For example, generating a report from legal documents might involve parsing the source files, extracting relevant evidence, and composing the final report.
We denote the subtask sequence formed up to execution stage $t$ as
\begin{equation}
    \mathcal{T}_{t}
    =
    (\tau_1,\ldots,\tau_t),
\end{equation}
where $\tau_t$ denotes the current subtask.
Based on the current subtask sequence and the retrieved candidate skills, the invocation process $\mathcal{I}$ determines the skill to be invoked:
\begin{equation}
    s_t^{\mathrm{inv}}
    =
    \mathcal{I}
    \left(
        \mathcal{T}_t,
        \mathcal{C}
    \right),
    \qquad
    s_t^{\mathrm{inv}}
    \in
    \mathcal{C}\cup\{\varnothing\},
\end{equation}
where $s_t^{\mathrm{inv}}=\varnothing$ indicates that the agent proceeds without invoking any candidate skill.

As discussed in the Introduction, Task--Skill Misalignment (TSM) might arise at both stages of skill identification.
We first characterize its manifestation in skill retrieval.

\begin{definition}[Retrieval-side Task--Skill Misalignment]
\label{def:retrieval_tsm}
Given a task $q$, let $s^{+}$ denote a useful skill for task execution and $s^{-}$ denote a less useful skill.
Retrieval-side TSM occurs when
\begin{equation}
    r(q,s^{+})
    \leq
    r(q,s^{-}),
\end{equation}
where $r(q,s)$ denotes the retrieval score between task $q$ and skill $s$.
\end{definition}

In this case, an execution-useful skill receives lower retrieval priority than a less useful one and might consequently be excluded from the candidate set $\mathcal{C}$.
Moreover, TSM might persist as follows even when the required skill has already been successfully retrieved.

\begin{definition}[Invocation-side Task--Skill Misalignment]
\label{def:invocation_tsm}
For a subtask $\tau_t$, let $s_t^{*}\in\mathcal{C}$ denote the skill required to accomplish it.
Invocation-side TSM occurs when the required skill is available but not invoked, \textit{i.e.},
\begin{equation}
    s_t^{*}\in\mathcal{C},
    \qquad
    s_t^{\mathrm{inv}}
    \neq
    s_t^{*}.
\end{equation}
\end{definition}

Based on the above formulation, TSM manifests as two complementary failures: execution-critical skills might be under-prioritized during retrieval or remain uninvoked during execution.
To address both, we instantiate the prospective reasoning introduced in the Introduction at the retrieval and invocation stages, respectively.
In the following, we elaborate on the corresponding designs.

\subsection{Prospective Skill Generation}
\label{sec:prospective_generation}

As discussed above, the task objective might not explicitly reveal the capabilities required for successful execution, making direct task--skill matching prone to overlooking execution-critical skills.
To remedy this, we introduce Prospective Skill Generation (PSG), which first uncovers the capability requirements underlying the task and then prospectively imagines how these capabilities could be realized.

Specifically, given a task $q$, we employ an off-the-shelf LLM $\mathcal{M}$ to infer a set of capability requirements $\mathcal{U}=\{u_j\}_{j=1}^{M}$.
Different from simply decomposing the explicit task description into finer-grained subqueries, PSG further reasons about the implicit capabilities required for task execution.
Accordingly, each $u_j$ characterizes an independent capability that requires specialized skill guidance, while trivial operations that could be readily handled by the agent itself are excluded.

Although capability inference could uncover the execution requirements underlying the task, reformulating the original task into capability-level descriptions might discard task-specific semantic information.
Such contextual information remains important for distinguishing skills that provide similar capabilities but target different application scenarios.
To preserve such information, PSG further extracts a set of semantic anchors $\mathcal{A}=\{a_l\}_{l=1}^{L}$ from the original task, including informative cues such as domain terminology, key objects, and task targets.
Accordingly, the capability inference process is formulated as
\begin{equation}
    \left(
        \mathcal{U},\mathcal{A}
    \right)
    =
    \mathcal{M}
    \left(
        q;\mathcal{P}_{\mathrm{PSG}}^{\mathrm{inf}}
    \right),
\end{equation}
where $\mathcal{P}_{\mathrm{PSG}}^{\mathrm{inf}}$ denotes the capability inference prompt.

Although $\mathcal{U}$ reveals the capabilities required for task execution, such concise capability descriptions still provide limited information about how these capabilities could be realized in practice.
As a result, a discrepancy remains between capability-level descriptions and real skills that provide detailed execution guidance.
To further bridge this discrepancy, PSG performs prospective reasoning to imagine the execution guidance that a useful skill would provide for each inferred capability.
Specifically, for each $u_j\in\mathcal{U}$, PSG generates a corresponding pseudo skill as
\begin{equation}
    \tilde{s}_j
    =
    \mathcal{M}
    \left(
        u_j,\mathcal{A};
        \mathcal{P}_{\mathrm{PSG}}^{\mathrm{gen}}
    \right),
    \qquad
    j=1,\ldots,M,
\end{equation}
where $\mathcal{P}_{\mathrm{PSG}}^{\mathrm{gen}}$ denotes the prospective generation prompt.
Accordingly, the generated pseudo skill set is denoted as $\widetilde{\mathcal{S}}=\{\tilde{s}_j\}_{j=1}^{M}$.
Rather than merely restating the inferred capability, each pseudo skill is formulated in a skill-like form that describes how the corresponding capability could be realized through concrete procedural guidance.

In this way, PSG transforms implicit capability requirements into execution-oriented pseudo skills while preserving task-specific semantics.
These pseudo skills thereby provide prospective retrieval cues that better align with the procedural knowledge encoded in real skills.

\subsection{Hybrid Skill Retrieval}
\label{sec:hybrid_retrieval}

Although PSG provides skill-like descriptions of the anticipated execution requirements, the generated pseudo skills are imagined rather than verified and thus should not be directly treated as executable skills.
Instead, we employ them as retrieval proxies and introduce Hybrid Skill Retrieval (HSR), which jointly considers the original task and generated pseudo skills to identify useful skills from the gallery.

To be specific, we first encode the task $q$, each pseudo skill $\tilde{s}_j\in\widetilde{\mathcal{S}}$, and each gallery skill $s_i\in\mathcal{S}$ using an embedding model $f(\cdot)$:
\begin{equation}
    \mathbf{z}_{q}=f(q), \qquad
    \tilde{\mathbf{z}}_{j}=f(\tilde{s}_{j}), \qquad
    \mathbf{z}_{i}=f(s_i).
\end{equation}
We then calculate the relevance between the original task and each gallery skill as
\begin{equation}
    r_i^{q}
    =
    \mathrm{sim}
    \left(
        \mathbf{z}_{q},
        \mathbf{z}_{i}
    \right),
\end{equation}
where $r_i^{q}$ measures the relevance of gallery skill $s_i$ to the original task.
Meanwhile, each pseudo skill is used as a retrieval proxy to identify gallery skills with similar capabilities and procedural guidance:
\begin{equation}
    r_{ij}^{p}
    =
    \mathrm{sim}
    \left(
        \tilde{\mathbf{z}}_{j},
        \mathbf{z}_{i}
    \right),
\end{equation}
where $r_{ij}^{p}$ measures the relevance between pseudo skill $\tilde{s}_j$ and gallery skill $s_i$.

To jointly account for task relevance and prospective execution information, we formulate the following fusion score:
\begin{equation}
    d_{ij}
    =
    \alpha r_i^{q}
    +
    r_{ij}^{p},
\end{equation}
where $\alpha$ controls the contribution of the original task relevance.
Unless otherwise specified, we set $\alpha=2$.
For each pseudo skill $\tilde{s}_j$, the top-$K_c$ gallery skills ranked by $d_{ij}$ are retained as the preliminary candidate set $\mathcal{B}_{j}$.

Although the above fusion incorporates both task and execution information, embedding-based matching might still underemphasize task-specific semantic cues preserved in the original task.
To further retain such information, HSR employs the semantic anchors $\mathcal{A}$ extracted by PSG for BM25-based reranking within each candidate set $\mathcal{B}_{j}$.
Let $a$ denote the lexical query constructed by concatenating the semantic anchors in $\mathcal{A}$.
For each candidate $s_i\in\mathcal{B}_{j}$, its anchor-based relevance is calculated as
\begin{equation}
    r_i^{a}
    =
    \mathrm{BM25}
    \left(
        a,s_i
    \right),
    \qquad
    s_i\in\mathcal{B}_{j}.
\end{equation}
Since the three relevance scores might have different numerical scales, we apply min-max normalization within each candidate set, yielding $\bar{r}_{i}^{q}$, $\bar{r}_{ij}^{p}$, and $\bar{r}_{i}^{a}$.
The final reranking score is then formulated as
\begin{equation}
    h_{ij}
    =
    \alpha \bar{r}_{i}^{q}
    +
    \bar{r}_{ij}^{p}
    +
    \bar{r}_{i}^{a}.
    \label{eq:fusion_results}
\end{equation}
Each pseudo skill therefore produces a capability-specific ranking $\mathcal{L}_{j}$ according to $h_{ij}$.
Since a complex task might involve multiple complementary capabilities, selecting skills from only one ranking could overlook part of the inferred capability requirements.
We therefore retain the top-$K_r$ skills from each ranking and organize them into a ranking matrix.
Specifically, let $\pi_j(v)$ denote the index of the $v$-th ranked gallery skill in $\mathcal{L}_{j}$, and define
\begin{equation}
    \mathbf{R}
    =
    \begin{bmatrix}
        s_{\pi_1(1)} & s_{\pi_2(1)} & \cdots & s_{\pi_M(1)} \\
        s_{\pi_1(2)} & s_{\pi_2(2)} & \cdots & s_{\pi_M(2)} \\
        \vdots       & \vdots       & \ddots & \vdots       \\
        s_{\pi_1(K_r)} & s_{\pi_2(K_r)} & \cdots & s_{\pi_M(K_r)}
    \end{bmatrix},
\end{equation}
where the $M$ columns correspond to the inferred capability requirements and the $K_r$ rows correspond to their ranking positions.
Starting from the first row, HSR scans $\mathbf{R}$ from left to right before proceeding to the next row, while skipping duplicate skills.
The process terminates once $K$ unique skills are collected, yielding the final candidate set
\begin{equation}
    \mathcal{C}
    =
    \left\{
        \hat{s}_{1},
        \hat{s}_{2},
        \ldots,
        \hat{s}_{K}
    \right\}.
\end{equation}
In this way, different capability requirements can contribute their highly ranked skills to the final candidate set, thereby improving the coverage of skills needed for subsequent execution.
\subsection{Capability-aware Skill Invocation}
\label{sec:capability_invocation}

Although HSR retrieves candidate skills $\mathcal{C}$ before execution, the current subtask may not explicitly reveal the capability required to accomplish its objective, causing an available execution-critical skill to remain uninvoked as detailed in the Introduction.
To address such invocation-side TSM, we introduce Capability-aware Skill Invocation (CSI), which makes the implicit capability requirement explicit and uses it to guide skill invocation during execution.

Following Sec.~\ref{sec:problem_formulation}, let $\mathcal{T}_t$ denote the subtask sequence formed up to execution stage $t$.
During execution, the agent progressively accomplishes the current subtask by selecting and performing concrete actions.
Before taking the next action, CSI reevaluate the current objective, infers the capability requirement $u_t$, and determines whether additional skill guidance is needed as follows:
\begin{equation}
    \left(
        u_t,
        s_t^{\mathrm{inv}}
    \right)
    =
    \mathcal{I}_{\mathrm{CSI}}
    \left(
        \mathcal{T}_t,
        \mathcal{C};
        \mathcal{P}_{\mathrm{CSI}},
        \mathcal{X}_n
    \right),
\end{equation}
where $\mathcal{P}_{\mathrm{CSI}}$ denotes the capability-aware invocation instruction and $\mathcal{X}_n$ denotes the runtime context introduced below.
If a candidate skill provides the required capability, CSI invokes it to support the next action; otherwise, $s_t^{\mathrm{inv}}=\varnothing$ and the agent proceeds without additional skill guidance.
Accordingly, CSI performs skill invocation through an explicit task objective--capability--skill reasoning process instead of directly matching the current subtask with candidate skills.
The detailed assessment format and prompts are provided in the supplementary materials.

During long-horizon execution, the capability-aware instruction may receive diminishing attention as the interaction context grows.
To maintain consistent capability assessment, CSI further injects a periodic runtime context after every $N$ executed actions:
\begin{equation}
    \mathcal{X}_n
    =
    \begin{cases}
        \mathcal{X}, & n \bmod N = 0,\\
        \varnothing, & \text{otherwise},
    \end{cases}
\end{equation}
where $n$ denotes the number of executed actions, $N$ denotes the injection interval and is fixed as $5$ for simplicity, and $\mathcal{X}$ denotes the runtime context that reinforces capability-aware invocation.
Such periodic context helps preserve the capability-aware decision process throughout long-horizon execution.
Thanks to the explicit capability reevaluation  before action, CSI could enables retrieved skills to be invoked when their guidance becomes necessary, thereby alleviating invocation-side TSM.
\section{Experiments}

In this section, we conduct extensive experiments on three representative agent skill benchmarks to evaluate the effectiveness of SkillDreamer in both skill retrieval and end-to-end task execution.
Due to space limitation, we present more experimental details and results in the Appendix.

\subsection{Experimental Setup}

\noindent\textbf{Benchmarks and Metrics.}
We evaluate SkillDreamer on three representative agent skill benchmarks, including SkillRet~\citep{skillret}, SkillUsage~\citep{skillusage}, and SkillNet-Gym~\citep{skillnet}.
SkillRet focuses on large-scale skill retrieval with 4,997 test queries and 6,660 candidate skills, while SkillUsage and SkillNet-Gym further support executable end-to-end evaluation.
SkillUsage contains 87 executable tasks and 34,396 candidate skills, of which 84 valid tasks are retained following its official protocol.
Following their official protocols, we report Recall@$K$ with $K\in\{5,10,15\}$ for skill retrieval, and \textit{Pass} and \textit{Load} for end-to-end task execution, measuring task completion and whether provided skills are loaded during execution, respectively.
More benchmark statistics and evaluation details are provided in the Appendix.

\noindent\textbf{Implementation Details.}
Unless otherwise specified, we employ Qwen3.6-27B~\citep{qwen36_27b} for PSG and Qwen3-Embedding-0.6B~\citep{qwen3embedding} as the default retriever.
PSG infers at most five capability requirements for each task, while HSR retains $K_c=20$ candidates for each inferred capability before reranking.
For end-to-end evaluation, we provide the top-five retrieved skills as candidate skills to GPT-5.5~\citep{openai_gpt55} and DeepSeek-V4-Flash~\citep{deepseek_v4}.
During execution, CSI is applied to determine whether a candidate skill should be invoked.
Additional implementation details are provided in the Appendix.

\subsection{Experiments on Skill Retrieval}

\begin{table*}[!t]
\caption{
Retrieval results (\%) on SkillUsage, SkillRet, and SkillNet-Gym.
SkillDreamer is applied plug-and-play to representative retrievers. The best results within each backbone are shown in \textbf{bold}.
}
\label{tab:retrieval_main}
\centering

{
\setlength{\tabcolsep}{3pt}
\renewcommand{\arraystretch}{0.82}

\begin{tabular*}{\textwidth}{
@{\extracolsep{\fill}}
l
c
rr
rr
rr
@{}
}
\toprule

\multirow{2}{*}{\textbf{Retriever}}
&
\multirow{2}{*}{\textbf{Variant}}
&
\multicolumn{2}{c}{\textbf{SkillUsage}}
&
\multicolumn{2}{c}{\textbf{SkillRet}}
&
\multicolumn{2}{c}{\textbf{SkillNet-Gym}}
\\

\cmidrule(lr){3-4}
\cmidrule(lr){5-6}
\cmidrule(lr){7-8}

&
&
\multicolumn{1}{c}{\textbf{R@5}}
&
\multicolumn{1}{c}{\textbf{R@10}}
&
\multicolumn{1}{c}{\textbf{R@5}}
&
\multicolumn{1}{c}{\textbf{R@10}}
&
\multicolumn{1}{c}{\textbf{R@5}}
&
\multicolumn{1}{c}{\textbf{R@10}}
\\

\midrule
\multicolumn{8}{>{\columncolor{gray!10}}c}{\textit{\textbf{Sparse Retrieval}}}
\\[2pt]
\\[-2pt]

BM25
& Original
& 49.06 & 55.71
& 49.66 & 56.17
& 48.90 & 60.64
\\

\multicolumn{8}{>{\columncolor{gray!10}}c}{\textit{\textbf{General Embedding Retrievers}}}
\\[2pt]
\\[-2pt]

BGE-large-en-v1.5
& Original
& 47.56 & 53.88
& 55.96 & 61.40
& 48.38 & 58.63
\\

E5-large-v2
& Original
& 46.42 & 48.29
& 51.51 & 57.42
& 43.11 & 51.01
\\

Qwen3-Embedding-8B
& Original
& 53.39 & 60.95
& 60.64 & 67.10
& 52.26 & 64.74
\\

\noalign{\vskip 3pt}
\cdashline{1-8}[2pt/3pt]
\noalign{\vskip 3pt}

\multirow[c]{3}{*}{Qwen3-Embedding-0.6B}
& Original
& 52.68 & 62.25
& 59.14 & 64.91
& 48.60 & 61.77
\\

&
\textbf{+ Ours}
& \textbf{60.42}
& \textbf{69.47}
& \textbf{66.93}
& \textbf{71.73}
& \textbf{60.92}
& \textbf{69.51}
\\

&
$\Delta$
& $+7.74$
& $+7.22$
& $+7.79$
& $+6.82$
& $+12.32$
& $+7.74$
\\

\multicolumn{8}{>{\columncolor{gray!10}}c}{\textit{\textbf{Skill-oriented Retrievers}}}
\\[2pt]
\\[-2pt]

\multirow[c]{3}{*}{SkillRouter-Embedding-0.6B}
& Original
& 46.55 & 52.29
& 71.01 & 76.00
& 39.62 & 48.56
\\

&
\textbf{+ Ours}
& \textbf{56.15}
& \textbf{61.69}
& \textbf{75.54}
& \textbf{81.00}
& \textbf{61.36}
& \textbf{68.43}
\\

&
$\Delta$
& $+9.60$
& $+9.40$
& $+4.53$
& $+5.00$
& $+21.74$
& $+19.87$
\\

\noalign{\vskip 3pt}
\cdashline{1-8}[2pt/3pt]
\noalign{\vskip 3pt}

\multirow[c]{3}{*}{SkillRet-Embedding-0.6B}
& Original
& 57.88 & 65.97
& 79.14 & 85.33
& 55.68 & 66.04
\\

&
\textbf{+ Ours}
& \textbf{60.80}
& \textbf{68.81}
& \textbf{79.91}
& \textbf{85.53}
& \textbf{63.55}
& \textbf{74.70}
\\

&
$\Delta$
& $+2.92$
& $+2.84$
& $+0.77$
& $+0.20$
& $+7.87$
& $+8.66$
\\

\noalign{\vskip 3pt}
\cdashline{1-8}[2pt/3pt]
\noalign{\vskip 3pt}

\multirow[c]{3}{*}{R3-Embedding}
& Original
& 55.50 & 61.17
& 81.82 & 87.64
& 54.70 & 66.79
\\

&
\textbf{+ Ours}
& \textbf{59.98}
& \textbf{67.51}
& \textbf{82.76}
& \textbf{87.81}
& \textbf{66.39}
& \textbf{75.59}
\\

&
$\Delta$
& $+4.48$
& $+6.34$
& $+0.94$
& $+0.17$
& $+11.69$
& $+8.80$
\\

\bottomrule
\end{tabular*}
}

\end{table*}

To comprehensively evaluate SkillDreamer on skill retrieval, we compare it with eight representative retrievers from three categories, namely,
i) sparse retrievers, including BM25~\citep{bm25};
ii) general embedding retrievers, including BGE-large-en-v1.5~\citep{bge}, E5-large-v2~\citep{e5}, and Qwen3-Embedding at different scales~\citep{qwen3embedding};
and iii) skill-oriented retrievers, including SkillRouter~\citep{skillrouter}, SkillRet~\citep{skillret}, and R3~\citep{r3}.
To examine whether prospective skill identification could mitigate TSM during retrieval across different retrieval paradigms, we further apply SkillDreamer to Qwen3-Embedding-0.6B and three skill-oriented retrievers in a plug-and-play manner.

From the results in Table~\ref{tab:retrieval_main}, one could have the following observations and conclusions.
i) \textbf{SkillDreamer consistently improves the retrieval performance across different benchmarks and retrievers.}
For example, applying SkillDreamer to Qwen3-Embedding-0.6B improves R@5 by 7.74, 7.79, and 12.32 percentage points on SkillUsage, SkillRet, and SkillNet-Gym, respectively.
Consistent improvements are also observed on three skill-oriented retrievers, demonstrating the effectiveness and broad compatibility of prospective skill identification in mitigating TSM during retrieval.
ii) \textbf{Stronger semantic matching does not always lead to better identification of execution-useful skills.}
Although SkillRouter is built upon Qwen3-Embedding and further trained with large-scale skill data, it underperforms the base model on both SkillUsage and SkillNet-Gym.
In contrast, SkillDreamer substantially improves SkillRouter on both benchmarks, indicating that strengthening task--skill semantic matching alone cannot bridge the discrepancy between objective-oriented tasks and execution-oriented skills, and is therefore insufficient to resolve TSM during retrieval.

\subsection{Experiments on End-to-End Task Execution}

\begin{table*}[!t]
\caption{
End-to-end task execution results (\%) on SkillUsage and SkillNet-Gym.
\textit{Pass} denotes the task completion rate, while
\textit{Load} denotes the proportion of execution trajectories in which
the agent loads at least one provided skill.
Best results among non-oracle settings are shown in \textbf{bold}.
}
\label{tab:skillusage_e2e}
\centering

{
\setlength{\tabcolsep}{0.5pt}
\renewcommand{\arraystretch}{0.82}
\setlength{\dashlinedash}{1.8pt}
\setlength{\dashlinegap}{1.4pt}

\begin{tabular}{
    @{}
    >{\raggedright\arraybackslash}p{2.85cm}
    *{8}{>{\centering\arraybackslash}p{1.35cm}}
    @{}
}
\toprule

\textbf{Benchmark}
& \multicolumn{4}{c}{\textbf{SkillUsage}}
& \multicolumn{4}{c}{\textbf{SkillNet-Gym}}
\\
\cmidrule{1-9}

\multirow{2}{*}{\textbf{Agent}}
& \multicolumn{2}{c}{\textbf{GPT-5.5}}
& \multicolumn{2}{c}{\makebox[0pt][c]{\textbf{DeepSeek-V4-Flash}}}
& \multicolumn{2}{c}{\textbf{GPT-5.5}}
& \multicolumn{2}{c}{\makebox[0pt][c]{\textbf{DeepSeek-V4-Flash}}}
\\

\cmidrule(l{1pt}r{1pt}){2-3}
\cmidrule(l{1pt}r{1pt}){4-5}
\cmidrule(l{1pt}r{1pt}){6-7}
\cmidrule(l{1pt}r{1pt}){8-9}

& \textbf{Pass} & \textbf{Load}
& \textbf{Pass} & \textbf{Load}
& \textbf{Pass} & \textbf{Load}
& \textbf{Pass} & \textbf{Load}
\\
\midrule

Ground-Truth Skills
& 59.0 & 97.6
& 46.3 & 72.6
& 66.1 & 100.0
& 61.0 & 40.7
\\

\noalign{\vskip 2pt}
\cdashline{1-9}[1.8pt/1.4pt]
\noalign{\vskip 2pt}

No Skills
& 46.8 & --
& 27.5 & --
& 69.5 & --
& 61.0 & --
\\

Direct Retrieval
& 48.1 & 92.9
& 37.1 & 51.2
& 69.5 & 93.2
& 59.3 & 30.5
\\


Ours
& \textbf{50.6} & \textbf{96.4}
& \textbf{41.4} & \textbf{77.4}
& \textbf{72.9} & \textbf{96.6}
& \textbf{66.1} & \textbf{47.5}
\\

\bottomrule
\end{tabular}
}
\end{table*}
In this section, we further evaluate SkillDreamer on end-to-end task execution using GPT-5.5 and DeepSeek-V4-Flash.
For comparison, we consider three settings:
i) \textit{Ground-Truth Skills} directly provides the annotated skills;
ii) \textit{No Skills} performs execution without external skill support;
and iii) \textit{Direct Retrieval} directly retrieves candidate skills using the original task.
For all skill-assisted settings, the agent is provided with five  candidate skills and autonomously decides whether to invoke the provided skills during execution.
As shown in Table~\ref{tab:skillusage_e2e}, even when ground-truth skills are directly provided, DeepSeek-V4-Flash achieves a Load of only 72.6\% on SkillUsage, indicating that skill availability alone does not guarantee successful invocation.
Compared with Direct Retrieval, the proposed SkillDreamer consistently improves both Pass and Load across different agents and benchmarks.
These results demonstrate that prospective skill identification could facilitate effective skill identification and consequently improve downstream task execution.

\subsection{Ablation and Analytic Studies}

In this section, we conduct ablation and analytic studies to further investigate the effectiveness of SkillDreamer.
Unless otherwise specified, all experiments are conducted on SkillUsage with Qwen3-Embedding-0.6B as the retriever.

\begin{table*}[t]
\caption{
Ablation studies of SkillDreamer for skill retrieval and invocation.
}
\label{tab:ablation}
\centering

\begin{minipage}[t]{0.48\textwidth}
\centering
\textbf{(a) Skill Retrieval}

\vspace{3pt}
\footnotesize
\setlength{\tabcolsep}{2.7pt}
\renewcommand{\arraystretch}{1.05}
\begin{tabular}{@{}lcccc@{}}
\toprule
\multirow{2}{*}{\textbf{Setting}}
& \multicolumn{2}{c}{\textbf{SkillRet}}
& \multicolumn{2}{c}{\textbf{SkillUsage}} \\
\cmidrule(lr){2-3}
\cmidrule(lr){4-5}
& \textbf{R@5} & \textbf{R@10}
& \textbf{R@5} & \textbf{R@10} \\
\midrule

Direct Retrieval
& 59.14 & 64.91
& 52.68 & 62.25 \\

Pseudo Skill w/o Cap.
& 49.56 & 55.15
& 48.51 & 58.69 \\

Pseudo-Skill Retrieval
& 55.56 & 62.60
& 48.85 & 59.00 \\

PSG + HSR
& \textbf{66.93} & \textbf{71.73}
& \textbf{60.42} & \textbf{69.47} \\

\bottomrule
\end{tabular}
\end{minipage}
\hfill
\begin{minipage}[t]{0.5\textwidth}
\centering
\textbf{(b) Skill Invocation}

\vspace{3pt}
\footnotesize
\setlength{\tabcolsep}{2.2pt}
\renewcommand{\arraystretch}{1.05}
\begin{tabular}{@{}lcccc@{}}
\toprule
\multirow{2}{*}{\textbf{Setting}}
& \multicolumn{2}{c}{\textbf{GPT-5.5}}
& \multicolumn{2}{c}{\shortstack{\textbf{DeepSeek-}\textbf{V4-Flash}}} \\
\cmidrule(lr){2-3}
\cmidrule(lr){4-5}
& \textbf{Pass} & \textbf{Load}
& \textbf{Pass} & \textbf{Load} \\
\midrule

No Skills
& 46.8 & --
& 27.5 & -- \\

Pseudo Skills
& 42.9 & 90.5
& 25.8 & \textbf{78.6} \\

PSG + HSR
& 49.7 & 95.2
& 38.5 & 54.8 \\

SkillDreamer
& \textbf{50.6} & \textbf{96.4}
& \textbf{41.4} & 77.4 \\

\bottomrule
\end{tabular}
\end{minipage}

\end{table*}

\noindent\textbf{Ablation Study.}
To investigate the contribution of different designs in SkillDreamer, we construct several variants for the retrieval and invocation stages, respectively.

For skill retrieval, Table~\ref{tab:ablation}(a) compares:
i) \textit{Direct Retrieval}, which retrieves skills using the original task;
ii) \textit{Pseudo Skill w/o Cap.}, which directly generates pseudo skills from the task for retrieval;
iii) \textit{Pseudo-Skill Retrieval}, which first infers required capabilities through PSG and then retrieves with the generated pseudo skills;
and iv) \textit{PSG + HSR}, which denotes the complete retrieval stage.
From the results, PSG + HSR achieves the best performance, demonstrating that explicitly reasoning about required capabilities could better bridge task objectives and execution-oriented skills for skill identification.

For skill invocation, Table~\ref{tab:ablation}(b) compares:
i) \textit{No Skills}, which performs the task without external skills;
ii) \textit{Pseudo Skills}, which directly provides the generated pseudo skills for execution;
iii) \textit{PSG + HSR}, which performs execution with the skills retrieved by our method;
and iv) \textit{SkillDreamer}, which further employs CSI for skill identification during execution.
Introducing CSI upon PSG + HSR consistently improves both Pass and Load, demonstrating that reasoning about capability requirements during execution could facilitate the identification and invocation of desired skills, thereby mitigating TSM at the invocation stage.




\begin{figure*}[t]
    \centering

    \begin{subfigure}[t]{0.32\textwidth}
        \centering
        \includegraphics[width=0.98\linewidth]{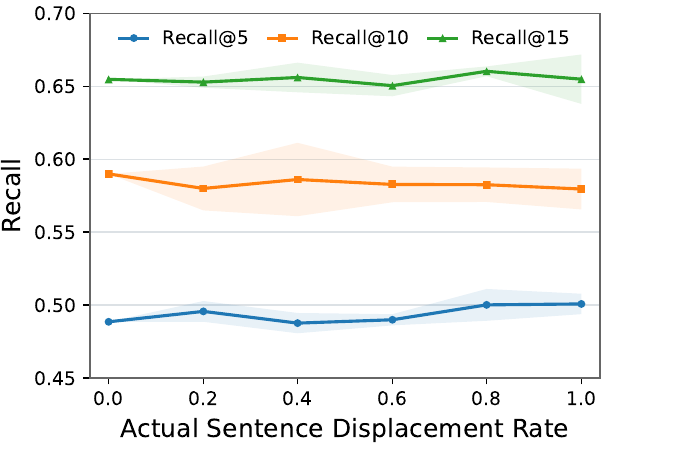}
        \caption{Pseudo-skill robustness.}
        \label{fig:pseudo_skill_sentence_shuffle}
    \end{subfigure}
    \hfill
    \begin{subfigure}[t]{0.32\textwidth}
        \centering
        \includegraphics[width=0.98\linewidth]{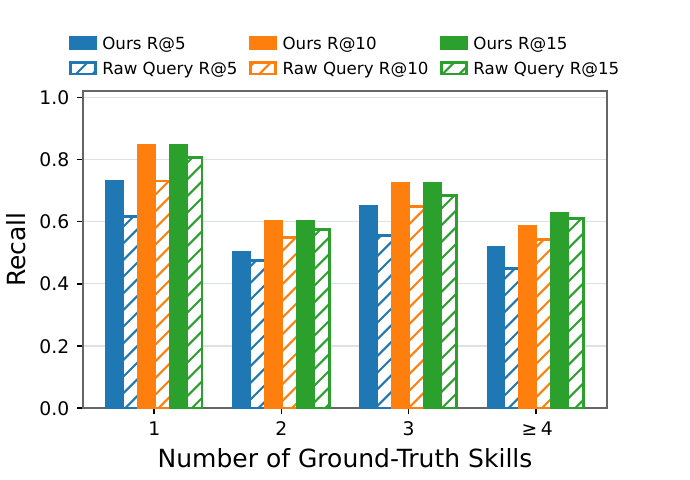}
        \caption{Number of ground-truth skills.}
        \label{fig:gt_skill_number}
    \end{subfigure}
    \hfill
    \begin{subfigure}[t]{0.32\textwidth}
        \centering
        \includegraphics[width=0.98\linewidth]{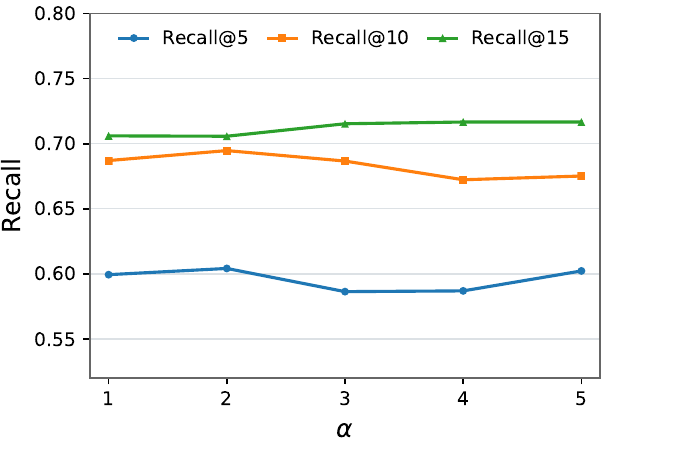}
        \caption{Sensitivity to $\alpha$.}
        \label{fig:alpha_sensitivity}
    \end{subfigure}

    \caption{
    Analytic studies of SkillDreamer.
    (a) Retrieval robustness under pseudo-skill sentence displacement.
    (b) Recall@$K$ under different numbers of ground-truth skills.
    (c) Sensitivity analysis of the hyperparameter $\alpha$ on SkillUsage.
    }
    \label{fig:analytic_studies}
\end{figure*}

\noindent\textbf{Robustness to Pseudo Skill Generation.}
To examine how the quality of generated pseudo skills affects SkillDreamer, we randomly reorder sentences within each pseudo skill at varying displacement rates under the ``Pseudo-Skill Retrieval" setting. 
As shown in Fig.~\ref{fig:pseudo_skill_sentence_shuffle}, retrieval performance remains stable even under substantial sentence displacement. 
This result suggests that an approximate description of how the required capabilities could be realized is sufficient to bridge task objectives and execution-oriented skills. 
In other words, SkillDreamer does not require generated pseudo skills to preserve precise procedural knowledge.

\noindent\textbf{Analysis on the Number of Ground-Truth Skills.}
We group SkillUsage tasks by the number of ground-truth skills to examine retrieval performance under varying capability requirements.
As shown in Fig.~\ref{fig:gt_skill_number}, SkillDreamer consistently outperforms Direct Retrieval across all groups, with clear gains retained as the number of required skills increases.
These results demonstrate that SkillDreamer could effectively identify the diverse skills required by tasks with multiple capability requirements.


\noindent\textbf{Parameter Analysis.}
In HSR, $\alpha$ balances the contribution of original-task relevance and prospective execution information during retrieval.
To investigate its sensitivity, we vary $\alpha$ from $1$ to $5$ on SkillUsage.
As shown in Fig.~\ref{fig:alpha_sensitivity}, SkillDreamer maintains stable performance across different values of $\alpha$, while $\alpha=2$ achieves competitive results on all metrics.
This observation indicates that HSR does not rely on a carefully tuned balance between task relevance and prospective execution information.

\section{Conclusion}

In this work, we reveal and study the Task--Skill Misalignment (TSM) problem in agent skill identification, where task descriptions mainly specify task objectives while skill documents provide the capabilities and procedural guidance required for execution.
We empirically demonstrate that such a discrepancy could hinder desirable skills from being correctly identified during both skill retrieval and invocation.
To address this problem, inspired by human prospective cognition, we propose SkillDreamer, a prospective skill identification framework that explicitly reasons about the capabilities required for execution to bridge task objectives and execution-oriented skills.
Extensive experiments on representative benchmarks not only demonstrate the necessity of mitigating TSM, but also verify the effectiveness and generalizability of SkillDreamer across diverse retrievers and LLM agents.
In the future, we plan to extend SkillDreamer to broader settings, including coding and GUI agents, and investigate how agents can more reliably follow the procedural guidance provided by identified skills.

\subsection*{AI use statement}
In this work, we used generative AI tools to assist with the analysis of experimental results and to improve the clarity, grammar, and academic presentation of the manuscript. Beyond the LLM-based methods and evaluation procedures described in this paper, generative AI was not used to generate experimental data. All AI-assisted analyses and edits were carefully reviewed, verified, and revised by the authors as needed. The authors made the final interpretations and conclusions and take full responsibility for the accuracy and integrity of the manuscript.

\putbib
\end{bibunit}

\clearpage
\appendix
\addtocontents{toc}{\protect\setcounter{tocdepth}{2}}
\setcounter{tocdepth}{2}
\tableofcontents
\clearpage
\renewcommand{\bibunitprefix}{supp.}
\begin{bibunit}
\section{Introduction}

In this supplementary material, we provide further implementation details and additional experimental results to demonstrate the effectiveness of SkillDreamer.

\section{Related Work}

In this section, we provide a brief review of three topics highly related to this work, including information retrieval, agent skill retrieval, and agent skill invocation.

\subsection{Information Retrieval}

Information retrieval aims to identify relevant information from large-scale corpora for a given query.
Existing approaches mainly estimate query--document relevance through lexical matching or learned semantic representations~\citep{bm25,dense}.
With the emergence of LLMs, recent studies further improve retrieval by enriching query-side information or reasoning over candidate relevance.
Based on how LLMs are incorporated into the retrieval process, existing approaches could be roughly divided into two categories:
i) pre-retrieval query enhancement methods~\citep{querywriting,query2doc,hyde}, which reformulate, expand, or decompose the original query, or synthesize auxiliary retrieval content to better express the underlying information need;
ii) post-retrieval relevance refinement methods~\citep{chatgpt,large,first}, which employ LLMs to reason over retrieved candidates and rerank them through pairwise or listwise comparisons.

Different from conventional information retrieval, agent skill identification is not merely concerned with finding information relevant to a query, but with identifying reusable capabilities and procedural guidance that could support task execution.
Although existing query enhancement methods enrich the information expressed by the original query, they primarily aim to better characterize the underlying information need without explicitly reasoning about the capabilities required for execution.
In contrast, SkillDreamer prospectively infers execution-required capabilities and imagines their realizations as skill-like procedural guidance, thereby bridging task objectives and execution-oriented skills.
Moreover, SkillDreamer extends such capability-aware identification beyond retrieval to skill invocation during execution.

\subsection{Agent Skill Retrieval}

Agent skill retrieval aims to identify useful skills from a large-scale skill gallery for a given task, thereby providing reusable procedural guidance for subsequent execution.
According to how task--skill association is improved, existing studies could be roughly divided into two categories:
i) model adaptation methods, which adapt general-purpose retrieval models to the skill domain using paired task descriptions and skills~\citep{skillret,r3};
ii) pipeline redesign methods, which either adopt multi-stage retrieval and reranking strategies~\citep{skillrouter,skillflow}, or exploit task and skill structures through query decomposition, dependency modeling, and structured organization~\citep{SKillRerank,gosskill,skillresolve}.

Different from existing studies that mainly improve the direct association between task descriptions and skill documents, our work revisits skill retrieval from the perspective of task execution.
Task descriptions mainly specify task objectives, whereas skill documents describe the capabilities and procedural guidance required for execution, giving rise to Task--Skill Misalignment (TSM) during retrieval.
As a result, execution-critical skills might be under-prioritized when their semantics are weakly aligned with the original task description.
To mitigate such misalignment, SkillDreamer explicitly reasons about the capabilities required for task execution and imagines how they could be realized as pseudo skills, thereby introducing prospective execution information to bridge task objectives and execution-oriented skills.

\subsection{Agent Skill Invocation}

Agent skill invocation aims to determine whether and which available skills should be invoked during task execution, thereby providing appropriate procedural guidance when needed.
Based on how invocation decisions are made, existing approaches could be roughly divided into two categories:
i) invocation policy optimization methods, which explicitly optimize whether or which skill to invoke through dedicated credit assignment or preference learning~\citep{skillgate,selskill};
ii) execution-conditioned invocation methods, which adapt skill invocation to evolving execution states using explicit interaction trajectories or latent agent representations~\citep{trace,gavel}.

Different from existing studies that directly derive skill invocation from the task or execution context, our work explicitly models the capabilities required by emerging subtasks.
During execution, emerging subtasks primarily specify what needs to be accomplished, whereas available skills describe the capabilities and procedural guidance they provide, leaving the capability requirements connecting the two implicit.
Such misalignment might cause desirable skills to remain uninvoked even when they have already been retrieved.
To address this issue, SkillDreamer continually reassesses the capabilities required by emerging subtasks and accordingly determines whether and which retrieved skills should be invoked.
\section{Additional Details of Implementation}
In this section, we provide additional implementation details of SkillDreamer. Specifically, we first present the prospective skill retrieval procedure in the form of pseudocode. We then introduce the details of the datasets, prompts, and baseline retrievers used in our experiments. Finally, we present the experimental settings and evaluation metrics to facilitate reproducibility.

\subsection{Details of the Algorithm}

To present the retrieval procedure clearly, we provide the algorithmic details of PSG and HSR in Algorithm~\ref{alg:skilldreamer}. The notation follows the method section of the main paper.

\begin{algorithm}[!htbp]
\caption{Prospective skill retrieval in SkillDreamer.}
\label{alg:skilldreamer}
\begin{algorithmic}[1]

\REQUIRE Task $q$, skill gallery $\mathcal{S}=\{s_i\}_{i=1}^{G}$, LLM $\mathcal{M}$, embedding model $f$, retrieval scales $K_c$, $K_r$, and $K$, fusion weight $\alpha$

\ENSURE Retrieved candidate skill set $\mathcal{C}$

\STATE Infer capability requirements and semantic anchors:
$(\mathcal{U},\mathcal{A})
\leftarrow
\mathcal{M}(q;\mathcal{P}^{\mathrm{inf}}_{\mathrm{PSG}})$

\STATE Let $M=|\mathcal{U}|$

\FOR{each capability $u_j\in\mathcal{U}$}

    \STATE Generate the corresponding pseudo skill:
    $\tilde{s}_j
    \leftarrow
    \mathcal{M}(u_j,\mathcal{A};\mathcal{P}^{\mathrm{gen}}_{\mathrm{PSG}})$

\ENDFOR

\STATE Let $\widetilde{\mathcal{S}}=\{\tilde{s}_j\}_{j=1}^{M}$

\STATE Encode the task, pseudo skills, and gallery skills:
$\mathbf{z}_q=f(q)$, $\tilde{\mathbf{z}}_j=f(\tilde{s}_j)$, and $\mathbf{z}_i=f(s_i)$

\STATE Construct the lexical query $a$ by concatenating the anchors in $\mathcal{A}$

\FOR{each pseudo skill $\tilde{s}_j\in\widetilde{\mathcal{S}}$}

    \FOR{each gallery skill $s_i\in\mathcal{S}$}

        \STATE Compute query--skill relevance:
        $r_i^q=\mathrm{sim}(\mathbf{z}_q,\mathbf{z}_i)$

        \STATE Compute pseudo-skill--skill relevance:
        $r_{ij}^{p}=\mathrm{sim}(\tilde{\mathbf{z}}_j,\mathbf{z}_i)$

        \STATE Compute hybrid retrieval score:
        $d_{ij}
        =
        \alpha r_i^q+r_{ij}^{p}$
        \label{line:hybrid_retrieval_score}

    \ENDFOR

    \STATE Retain the top-$K_c$ skills ranked by $d_{ij}$ as the preliminary candidate set $\mathcal{B}_j$
    \label{line:retrieve_candidates}

    \FOR{each candidate skill $s_i\in\mathcal{B}_j$}

        \STATE Compute anchor relevance:
        $r_i^a=\mathrm{BM25}(a,s_i)$

    \ENDFOR

    \STATE Apply min--max normalization to the three relevance scores in the current candidate set to obtain $\bar{r}_i^q$, $\bar{r}_{ij}^{p}$, and $\bar{r}_i^a$

    \FOR{each candidate skill $s_i\in\mathcal{B}_j$}
    
        \STATE Compute the final reranking score:
        $h_{ij}
        =
        \alpha\bar{r}_i^q
        +
        \bar{r}_{ij}^{p}
        +
        \bar{r}_i^a$
        \label{line:reranking_score}
    
    \ENDFOR

    \STATE Rank $\mathcal{B}_j$ by $h_{ij}$ to obtain $\mathcal{L}_j$ and retain its top-$K_r$ skills
    \label{line:rerank_candidates}

\ENDFOR

\STATE Let $\pi_j(v)$ be the index of the $v$-th ranked skill in $\mathcal{L}_j$

\STATE Construct the ranking matrix:
$\mathbf{R}\leftarrow
\begin{bmatrix}
s_{\pi_1(1)} & \cdots & s_{\pi_M(1)} \\
\vdots & \ddots & \vdots \\
s_{\pi_1(K_r)} & \cdots & s_{\pi_M(K_r)}
\end{bmatrix}$

\STATE Initialize $\mathcal{C}\leftarrow\varnothing$

\STATE Scan $\mathbf{R}$ row by row from left to right, appending each unseen skill to $\mathcal{C}$ until $|\mathcal{C}|=K$
\label{line:select_topk}

\RETURN $\mathcal{C}$

\end{algorithmic}
\end{algorithm}
\FloatBarrier

\subsection{Details of the Datasets}

We evaluate the proposed method SkillDreamer on three skill-retrieval datasets: SkillRet~\citep{skillret}, SkillUsage~\citep{skillusage}, and SkillNet-Gym~\citep{skillnet}.
\begin{itemize}
    \item SkillRet is a large-scale benchmark constructed from public skills collected from open-source communities. It generates natural language queries from sampled skills using an LLM, with the sampled skills serving as the corresponding retrieval targets.
    \item SkillUsage is a benchmark built upon SkillsBench~\citep{li2026skillsbench}. It contains executable tasks and their corresponding skills, which are integrated into a large-scale library of real-world skills for skill retrieval evaluation. 
    \item SkillNet-Gym is the execution benchmark introduced by SkillNet~\citep{skillnet}. We use its tasks and ground-truth skills but construct a different retrieval gallery by merging them with the SkillUsage gallery and deduplicating the normalized \texttt{SKILL.md} content.
\end{itemize}

Detailed statistics and the numbers of tasks used in the retrieval and end-to-end execution evaluations are summarized in Table~\ref{tab:dataset_statistics}.

\begin{table}[H]
\caption{Statistics of the datasets used in our experiments. The SkillNet-Gym gallery combines its 162 unique ground-truth skills with the SkillUsage gallery and contains 34,522 skills after normalization and deduplication.}
\label{tab:dataset_statistics}
\centering
\setlength{\tabcolsep}{5pt}
\begin{tabular}{@{}lccccc@{}}
\toprule
Dataset & Split & \shortstack{Retrieval\\Tasks} & \shortstack{End-to-End\\Tasks} & Skills & GT Skills \\
\midrule
SkillRet
& Test & 4,997 & -- & 6,660 & 1--3 \\
\midrule
SkillUsage
& Test & 87 & 84 & $\sim$34K & Varies \\
\midrule
SkillNet-Gym
& Test & 159 & 59 & 34,522 & Varies \\
\bottomrule
\end{tabular}
\end{table}

As shown in Table~\ref{tab:dataset_statistics}, the retrieval evaluation on SkillRet uses its complete test split of 4,997 tasks. For SkillUsage, retrieval is evaluated on all 87 tasks, while end-to-end execution is evaluated on 84 tasks. For SkillNet-Gym, retrieval is evaluated on all 159 tasks, while end-to-end execution uses the first 59 tasks in the benchmark ordering. For both SkillUsage and SkillNet-Gym, we use the same two agent configurations in the end-to-end evaluation: GPT-5.5 with Codex and DeepSeek-V4-Flash with QCoder. All results reported in the main paper are computed over their corresponding evaluation subsets.

\subsection{Details of the Prompts}

To facilitate reproducibility, we provide the prompts used by the Prospective Skill Generation (PSG) and Capability-aware Skill Invocation (CSI) modules described in the main paper. Table~\ref{tab:ci_prompt} details the capability-inference prompt $\mathcal{P}^{\mathrm{inf}}_{\mathrm{PSG}}$, which jointly infers independent capability requirements $\mathcal{U}$ and task-specific semantic anchors $\mathcal{A}$ from the original task. Table~\ref{tab:psg_prompt} shows the prospective-generation prompt $\mathcal{P}^{\mathrm{gen}}_{\mathrm{PSG}}$, which takes each inferred capability and the semantic anchors as input and generates a corresponding skill-like pseudo skill with concrete procedural guidance. Table~\ref{tab:csi_prompt} provides the capability-aware invocation instruction $\mathcal{P}_{\mathrm{CSI}}$, which infers the capability required by the current objective and determines whether an appropriate candidate skill should be invoked. Table~\ref{tab:runtime_checkpoint_prompt} gives the periodic runtime checkpoint $\mathcal{X}_n$, which is injected after every fixed number of execution steps to trigger another capability-gap check while accounting for the remaining execution budget.




\newcommand{\CSIPromptTable}{%
\noindent\begin{minipage}{\textwidth}
\captionsetup{hypcap=false}
\captionof{table}{Capability-aware invocation instruction $\mathcal{P}_{\mathrm{CSI}}$ used by the CSI module.}
\label{tab:csi_prompt}
\end{minipage}

\begin{tcolorbox}[
    enhanced,
    breakable,
    width=\textwidth,
    colback=gray!7,
    colframe=gray!65,
    boxrule=0.8pt,
    arc=2.5mm,
    outer arc=2.5mm,
    left=6pt,
    right=6pt,
    top=2pt,
    bottom=2pt,
    boxsep=0pt,
    before skip=0pt,
    after skip=0pt
]
\small
\raggedright

\textbf{Capability-aware Skill Invocation.}
Complete the task with the available tools and the fixed set of candidate skills exposed by the runtime. Do not search for, install, or create additional skills.

\vspace{0.1em}
\textbf{Inputs.}
Use the sub-task sequence formed so far, the metadata of all retrieved candidate skills, and the runtime context supplied at the current checkpoint.

\vspace{0.1em}
\textbf{Goal.}
Before the next action, reassess the current objective, infer the capability required to accomplish it, and determine whether an appropriate candidate skill should be invoked. If no candidate skill provides the required capability, proceed without additional skill guidance.

\vspace{0.1em}
\textbf{When to Check.}
Run a check whenever the runtime issues a \texttt{RUNTIME CAPABILITY CHECKPOINT}. Also check when the required capability may have changed, the next step is unclear, an approach fails, attempts repeat, or new evidence makes the current skill match uncertain.

\vspace{0.1em}
\textbf{Check and Match.}
Before any tool call, begin the response with exactly five concise lines:

\vspace{0.05em}
{\ttfamily
SUBGOAL: the immediate subgoal\\
EVIDENCE: relevant evidence and unresolved constraints\\
CAPABILITY: the capability needed to resolve the most important unresolved issue\\
DECISION: MATCH, PARTIAL\_MATCH, NO\_MATCH, or NO\_CAPABILITY\_GAP; include the selected skill when applicable\\
NEXT: one bounded action and its observable success check
}

\vspace{0.05em}
Infer \texttt{CAPABILITY} from the unresolved objective, requirement, failure, or constraint, not from the tool action or method already planned. Compare it with all visible candidate-skill metadata before choosing \texttt{NO\_CAPABILITY\_GAP}.

\begin{itemize}
    \setlength{\itemsep}{0pt}
    \setlength{\topsep}{1pt}
    \setlength{\parsep}{0pt}
    \setlength{\parskip}{0pt}
    \item Use \texttt{MATCH} when a skill covers the required capability, \texttt{PARTIAL\_MATCH} when it covers a necessary part, and \texttt{NO\_MATCH} when none does.
    \item Prefer the most specific skill that addresses the unresolved issue. Do not select a skill merely because it is the closest candidate.
    \item Use \texttt{NO\_CAPABILITY\_GAP} only when the available evidence supports the current method, it satisfies the known constraints, and no unresolved failure or issue could invalidate \texttt{NEXT}.
\end{itemize}

\textbf{Act and Verify.}
For \texttt{MATCH} or \texttt{PARTIAL\_MATCH}, load the selected skill before any other tool action. Follow its applicable procedure and use its available implementation unless observed evidence shows that it is unsuitable. Load a different skill only when new evidence changes the match.

For \texttt{NO\_CAPABILITY\_GAP}, perform \texttt{NEXT}. For \texttt{NO\_MATCH}, decompose the subgoal or make one small, materially different attempt with the permitted tools; continue independent work and report any remaining blocker.

Execute \texttt{NEXT} and verify its stated success check. If it fails or conflicts with known evidence, run another check before continuing. Do not repeat a failed approach without new evidence or a different method.

Before completion, inspect the persisted result itself and validate its structure, content, and required outputs against the task's observable requirements.

\end{tcolorbox}
}

\begin{table}[!htbp]
\caption{Capability-inference prompt $\mathcal{P}^{\mathrm{inf}}_{\mathrm{PSG}}$ used by the PSG module.}
\label{tab:ci_prompt}
\centering

\begin{tcolorbox}[
    enhanced,
    width=\textwidth,
    colback=gray!7,
    colframe=gray!65,
    boxrule=0.8pt,
    arc=2.5mm,
    outer arc=2.5mm,
    left=6pt,
    right=6pt,
    top=2pt,
    bottom=2pt,
    boxsep=0pt,
    before skip=0pt,
    after skip=0pt
]
\small
\raggedright

\textbf{Role.}
You are an expert in decomposing tasks into capability requirements for skill retrieval.

\vspace{0.05em}
\textbf{Raw Task.} \texttt{\{query\}}

\vspace{0.05em}
\textbf{Goal.}
Infer the independent capabilities required to complete the raw task and extract task-specific semantic anchors, including relevant domain terminology, key objects, and task targets. Each capability will later be transformed into a skill-like retrieval proxy.

\vspace{0.15em}
\textbf{Decomposition Principles.}

\begin{itemize}
    \setlength{\itemsep}{0pt}
    \setlength{\topsep}{1pt}
    \setlength{\parsep}{0pt}
    \setlength{\parskip}{0pt}
    \item Decompose the task by independent capability requirements, rather than execution steps.
    \item Preserve relevant technical domains, methods, tools, and
    constraints from the raw task.
    \item Specialized artifact processing may be treated as an independent capability when it requires specific technical expertise.
    \item Do not invent capabilities, tools, methods, or requirements unsupported by the raw task.
\end{itemize}

\textbf{Output Requirements.}

Return JSON only. Generate one to five capability requirements and a concise list of semantic anchors grounded in the raw task. Each capability should be independent, self-contained, and useful for specialized skill retrieval.

\vspace{0.05em}
\textbf{Output Format.}

{\ttfamily
\{\\
\quad ``capabilities": [\\
\qquad \{\\
\qquad\quad ``id": ``u1",\\
\qquad\quad ``name": ``Capability name",\\
\qquad\quad ``requirement": ``Self-contained capability requirement".\\
\qquad \}\\
\quad ],\\
\quad ``semantic\_anchors": [\\
\qquad ``Domain term, key object, or task target from the raw task".\\
\quad ]\\
\}
}

\end{tcolorbox}

\end{table}
\begin{table}[!htbp]
\caption{Prospective-generation prompt $\mathcal{P}^{\mathrm{gen}}_{\mathrm{PSG}}$ used by the PSG module.}
\label{tab:psg_prompt}
\centering

\begin{tcolorbox}[
    enhanced,
    width=\textwidth,
    colback=gray!7,
    colframe=gray!65,
    boxrule=0.8pt,
    arc=2.5mm,
    outer arc=2.5mm,
    left=6pt,
    right=6pt,
    top=2pt,
    bottom=2pt,
    boxsep=0pt,
    before skip=0pt,
    after skip=0pt
]
\small
\raggedright

\textbf{Role.}
Write one concise, retrieval-oriented skill for a single capability.

\vspace{0.05em}
\textbf{Inputs.} \texttt{\{capability\_requirement\}} and \texttt{\{semantic\_anchors\}}

\vspace{0.05em}
\textbf{Goal.}
Generate one hypothetical \texttt{SKILL.md} that describes how the inferred capability could be realized through concrete procedural guidance. Preserve the relevant semantic anchors so that the pseudo skill remains specific to the original task context.

\vspace{0.3em}
\textbf{Generation Principles.}

\begin{itemize}
    \setlength{\itemsep}{0pt}
    \setlength{\topsep}{1pt}
    \setlength{\parsep}{0pt}
    \setlength{\parskip}{0pt}
    \item Teach how to realize the capability instead of merely restating its description.
    \item Incorporate concrete domain knowledge, such as methods, algorithms, APIs, workflows, failure handling, or validation.
    \item Preserve capability-defining technical terms and constraints.
    \item Do not invent unsupported facts or broaden the skill into adjacent capabilities.
    \item Produce a concise, reusable, and standalone skill using only capability-specific sections.
\end{itemize}

\textbf{Output Requirements.}

Return only the complete \texttt{SKILL.md} content. The generated skill must follow the specified metadata structure and contain a concise skill name, trigger-oriented description, and a human-readable title.

\vspace{0.05em}
\textbf{Output Format.}

{\ttfamily
---\\
name: short-lowercase-kebab-case-name\\
description: ``Concise, trigger-oriented description of the capability.''\\
---\\
\# Human-Readable Skill Title\\
Write the task-adaptive skill body.
}

\end{tcolorbox}

\end{table}
\FloatBarrier

\CSIPromptTable

\FloatBarrier

\begin{table}[!htbp]
\caption{Periodic runtime capability-checkpoint prompt $\mathcal{X}_n$ used to trigger the CSI module.}
\label{tab:runtime_checkpoint_prompt}
\centering

\begin{tcolorbox}[
    enhanced,
    width=\textwidth,
    colback=gray!7,
    colframe=gray!65,
    boxrule=0.8pt,
    arc=2.5mm,
    outer arc=2.5mm,
    left=6pt,
    right=6pt,
    top=2pt,
    bottom=2pt,
    boxsep=0pt,
    before skip=0pt,
    after skip=0pt
]
\small
\raggedright
\ttfamily
RUNTIME CAPABILITY CHECKPOINT (periodic): \{completedSteps\} execution steps have completed. Agent execution budget: \{budgetSeconds\} seconds; elapsed: \{elapsedSeconds\} seconds; estimated remaining: \{remainingSeconds\} seconds. Before taking another action, apply the Runtime Capability-Gap Recovery protocol in the task instructions. Infer the capability needed for the immediate subgoal and compare it with the visible skill metadata. Record NO\_CAPABILITY\_GAP, MATCH, PARTIAL\_MATCH, or NO\_MATCH, then act on the result. Choose a NEXT action that can be completed and verified within the remaining budget while leaving time to recover from failure.
\end{tcolorbox}

\end{table}
\FloatBarrier

\subsection{Details of the Baselines}




We consider three categories of retrieval baselines in our experiments. 
The details of the corresponding baseline retrievers are provided as follows:

\begin{itemize}
    \item Sparse Retrieval. BM25~\citep{bm25} is adopted as the representative sparse retriever, ranking skills through term-level matching between the query and skill documents.

    \item General Embedding Retrieval. Qwen3-Embedding~\citep{qwen3embedding} performs instruction-aware dense semantic matching, while BGE-large-en-v1.5~\citep{bge} combines RetroMAE pretraining with large-scale contrastive learning, and E5-large-v2~\citep{e5} learns general-purpose representations from weakly supervised text pairs followed by InfoNCE-based contrastive training.

    \item Skill-oriented Retrieval. SkillRouter-Embedding-0.6B~\citep{skillrouter}, SkillRet-Embedding-0.6B~\citep{skillret}, and R3-Embedding~\citep{r3} are all optimized with LLM-generated query--skill supervision, but differ in their training strategies. SkillRouter adopts in-batch InfoNCE with semantic, lexical, and category-based hard negatives, whereas SkillRet uses MultipleNegativesRankingLoss with in-batch negatives. Beyond pairwise relevance, R3-Embedding is trained on queries associated with one or more relevant skills and adopts multi-positive InfoNCE with a sibling-reward term to preserve multi-skill relevance.
\end{itemize}



    




\subsection{Details of the Experimental Settings}

To facilitate reproducibility, we provide the detailed experimental settings of SkillDreamer. As shown in Algorithm~\ref{alg:skilldreamer}, for each inferred capability $u_j$, $K_c$ denotes the number of skills retained in the preliminary candidate set $\mathcal{B}_j$ according to the hybrid score $d_{ij}$ (Line~\ref{line:retrieve_candidates}), while $K_r$ denotes the number of skills retained from the reranked list $\mathcal{L}_j$ (Line~\ref{line:rerank_candidates}). The resulting capability-specific rankings are merged in round-robin order with duplicates removed until $K$ unique skills are collected in the final candidate set $\mathcal{C}$ (Line~\ref{line:select_topk}). Unless otherwise specified, we set $K_c=20$, $K_r=20$, and $K=15$.

Beyond the retrieval scales, we further specify the fusion weights for different embedding backbones. Algorithm~\ref{alg:skilldreamer} follows the main-paper formulation, $d_{ij}=\alpha r_i^q+r_{ij}^{p}$ and $h_{ij}=\alpha\bar r_i^q+\bar r_{ij}^{p}+\bar r_i^a$. To state the model-specific implementation in a unified form, we write the same scores as
\begin{equation}
    d_{ij}
    =
    \alpha r_i^q+\beta r_{ij}^{p},
\label{eq:hybrid_score}
\end{equation}
\begin{equation}
    h_{ij}
    =
    \alpha\bar r_i^q
    +\beta\bar r_{ij}^{p}
    +\gamma\bar r_i^a.
\label{eq:reranking_score}
\end{equation}
The main-paper formulation is recovered when $\beta=\gamma=1$. By default, we set $\alpha=2$, $\beta=1$, and $\gamma=1$. For SkillRet-Embedding-0.6B and R3-Embedding on the SkillRet benchmark, we use $\alpha=4$, $\beta=1$, and $\gamma=0.25$.
Notably, this alternative configuration is used exclusively to evaluate the applicability of SkillDreamer to SkillRet-Embedding-0.6B and R3-Embedding on SkillRet, while all other experiments retain the fixed default configuration.



\subsection{Details of the Evaluation Metrics}

We evaluate SkillDreamer from two complementary perspectives: skill retrieval and end-to-end task execution. For each query $q$, let $G_q$ denote its ground-truth skill set and $\mathcal{C}_q$ denote the final candidate set containing the top-$K$ retrieved skills. Using these definitions, we evaluate retrieval performance with the following three metrics:

\begin{itemize}
    \item Recall@K measures the proportion of ground-truth
    skills retrieved within the top-$K$ results:
    \begin{equation}
        \mathrm{Recall}_q@K =
        \frac{|\mathcal{C}_q \cap G_q|}{|G_q|}.
    \end{equation}

    \item NDCG@K further considers the ranking positions of retrieved ground-truth skills, assigning larger gains to relevant skills appearing at higher ranks:
    \begin{equation}
        \mathrm{NDCG}_q@K =
        \frac{
        \sum_{i=1}^{K}
        \frac{\mathrm{rel}_i}{\log_2(i+1)}
        }{
        \sum_{i=1}^{\min(|G_q|,K)}
        \frac{1}{\log_2(i+1)}
        },
    \end{equation}
    where $\mathrm{rel}_i=1$ if the skill at rank $i$ belongs to $G_q$, and $0$ otherwise.

    \item Completeness@K is a stricter metric that evaluates
    whether all ground-truth skills are retrieved within the top-$K$ results:
    \begin{equation}
        \mathrm{Completeness}_q@K =
        \mathbb{I}\!\left[G_q \subseteq \mathcal{C}_q\right].
    \end{equation}
\end{itemize}

We further evaluate end-to-end task execution using the following metrics:

\begin{itemize}
    \item Pass Rate measures task-execution performance by averaging the verifier rewards over all tasks:
    \begin{equation}
        \mathrm{PassRate} =
        \frac{1}{N}
        \sum_{i=1}^{N} r_i,
    \end{equation}
    where $r_i$ denotes the verifier reward for task $i$, with missing or invalid rewards treated as zero.

    \item Load Rate measures the proportion of tasks for which the agent loads at least one provided skill during execution:
    \begin{equation}
        \mathrm{LoadRate} =
        \frac{1}{N}
        \sum_{i=1}^{N}
        \mathbb{I}\!\left[|U_i|>0\right],
    \end{equation}
    where $U_i$ denotes the set of provided skills loaded by the agent for task $i$.
\end{itemize}

\section{Additional Experiments}

In this section, we present additional experimental results to further demonstrate the effectiveness of SkillDreamer. Specifically, we first provide additional end-to-end execution results to examine the impact of improved retrieval on downstream task performance. We then report complementary retrieval results and evaluate its generalization across SRA-Bench~\citep{su2026skill} and diverse retriever backbones. Finally, we provide additional ablation results and further investigate the effect of the number of ground-truth skills on SkillRet.

\subsection{Unified Prospective Retrieval and Execution}

We further investigate a unified setting in which Qwen3.6-27B performs both prospective capability reasoning and task execution using its default QCoder harness. We compare SkillDreamer with \textit{Raw Query Retrieval}, which retrieves skills directly from the original task query. As shown in Table~\ref{tab:e2e_qwen36}, SkillDreamer improves the Pass rate from 31.3\% to 37.6\% ($+6.3$ points), while the Load rate remains unchanged at 59.5\%. The gain without more frequent skill loading indicates that prospective reasoning retrieves complementary skills that better support task completion rather than merely increasing skill usage.

\begin{table}[H]
\caption{End-to-end execution results (\%) on SkillUsage using Qwen3.6-27B in the QCoder harness. Best values are in \textbf{bold}.}
\label{tab:e2e_qwen36}
\centering
\setlength{\tabcolsep}{14pt}
\begin{tabular}{@{}lcc@{}}
\toprule
\textbf{Skill Setting} & \textbf{Pass} & \textbf{Load} \\
\midrule
Raw Query Retrieval
& 31.3 & \textbf{59.5} \\

\textbf{SkillDreamer}
& \textbf{37.6} & \textbf{59.5} \\

$\Delta$
& +6.3 & +0.0 \\
\bottomrule
\end{tabular}
\end{table}

\subsection{Additional Retrieval Results}

Tables~\ref{tab:retrieval_skillusage_full}, \ref{tab:retrieval_skillret_full}, and~\ref{tab:retrieval_skillnet_full} report complete Recall@$K$, NDCG@$K$, and Completeness@$K$ results on SkillUsage, SkillRet, and SkillNet-Gym. With Qwen3-Embedding-0.6B, SkillDreamer improves R@5 by $7.74$, $7.79$, and $12.32$ points; NDCG@5 by $4.79$, $4.68$, and $10.68$ points; and Completeness@5 by $6.90$, $9.76$, and $11.95$ points, respectively, demonstrating consistent gains in both ranking quality and complete skill-set recovery.

\begingroup
\setlength{\intextsep}{7pt}

\begin{table}[H]
\caption{
Complete SkillUsage retrieval results (\%); per-backbone best results are in \textbf{bold}.
}
\label{tab:retrieval_skillusage_full}
\centering

{
\small
\setlength{\tabcolsep}{0.35pt}
\renewcommand{\arraystretch}{0.88}
\setlength{\dashlinedash}{1.6pt}
\setlength{\dashlinegap}{1.2pt}
\setlength{\arrayrulewidth}{0.4pt}

\begin{tabular}{
    @{}
    >{\raggedright\arraybackslash}p{3.75cm}
    >{\centering\arraybackslash}p{1.10cm}
    *{9}{>{\centering\arraybackslash}p{\dimexpr(\textwidth-4.85cm-20\tabcolsep)/9\relax}}
    @{}
}
\toprule

\multirow{2}{*}{\textbf{Retriever}}
& \multirow{2}{*}{\textbf{Setting}}
& \multicolumn{3}{c}{\textbf{Recall}}
& \multicolumn{3}{c}{\textbf{NDCG}}
& \multicolumn{3}{c}{\textbf{Comp.}}
\\
\cmidrule(lr){3-5}
\cmidrule(lr){6-8}
\cmidrule(lr){9-11}
& & \textbf{@5} & \textbf{@10} & \textbf{@15}
& \textbf{@5} & \textbf{@10} & \textbf{@15}
& \textbf{@5} & \textbf{@10} & \textbf{@15}
\\

\midrule
\multicolumn{11}{c}{\textbf{Sparse Retrieval}}
\\
\cmidrule{1-11}

BM25
& Original
& 49.06 & 55.71 & 61.59
& 48.34 & 50.86 & 52.79
& 33.33 & 40.23 & 45.98
\\

\midrule
\multicolumn{11}{c}{\textbf{General Embedding Retrievers}}
\\
\cmidrule{1-11}

BGE-large-en-v1.5
& Original
& 47.56 & 53.88 & 56.84
& 48.24 & 50.92 & 51.98
& 32.18 & 35.63 & 39.08
\\

E5-large-v2
& Original
& 46.42 & 48.29 & 50.48
& 46.26 & 46.93 & 47.74
& 32.18 & 32.18 & 33.33
\\

Qwen3-Embedding-8B
& Original
& 53.39 & 60.95 & 69.72
& 54.77 & 57.91 & 60.78
& 32.18 & 41.38 & 50.57
\\

\multirow[c]{3}{*}{Qwen3-Embedding-0.6B}
& Original
& 52.68 & 62.25 & 67.66
& 52.62 & 56.51 & 58.36
& 35.63 & 43.68 & 48.28
\\

& \textbf{+ Ours}
& \textbf{60.42} & \textbf{69.47} & \textbf{70.57}
& \textbf{57.41} & \textbf{60.83} & \textbf{61.33}
& \textbf{42.53} & \textbf{51.72} & \textbf{52.87}
\\

& $\Delta$
& $+7.74$ & $+7.22$ & $+2.91$
& $+4.79$ & $+4.32$ & $+2.97$
& $+6.90$ & $+8.04$ & $+4.59$
\\

\midrule
\multicolumn{11}{c}{\textbf{Skill-oriented Retrievers}}
\\
\cmidrule{1-11}

\multirow[c]{3}{*}{SkillRouter-Embedding-0.6B}
& Original
& 46.55 & 52.29 & 55.70
& 47.32 & 49.78 & 51.00
& 29.89 & 34.48 & 35.63
\\

& \textbf{+ Ours}
& \textbf{56.15} & \textbf{61.69} & \textbf{65.21}
& \textbf{55.57} & \textbf{57.80} & \textbf{58.97}
& \textbf{33.33} & \textbf{42.53} & \textbf{45.98}
\\

& $\Delta$
& $+9.60$ & $+9.40$ & $+9.51$
& $+8.25$ & $+8.02$ & $+7.97$
& $+3.44$ & $+8.05$ & $+10.35$
\\

\noalign{\vskip 2pt}
\cdashline{1-11}[1.8pt/1.4pt]
\noalign{\vskip 2pt}

\multirow[c]{3}{*}{SkillRet-Embedding-0.6B}
& Original
& 57.88 & 65.97 & 69.49
& 57.08 & 60.42 & 61.56
& 40.23 & 50.57 & 55.17
\\

& \textbf{+ Ours}
& \textbf{60.80} & \textbf{68.81} & \textbf{74.92}
& \textbf{59.39} & \textbf{62.59} & \textbf{64.55}
& \textbf{42.53} & \textbf{54.02} & \textbf{58.62}
\\

& $\Delta$
& $+2.92$ & $+2.84$ & $+5.43$
& $+2.31$ & $+2.17$ & $+2.99$
& $+2.30$ & $+3.45$ & $+3.45$
\\

\noalign{\vskip 2pt}
\cdashline{1-11}[1.8pt/1.4pt]
\noalign{\vskip 2pt}

\multirow[c]{3}{*}{R3-Embedding}
& Original
& 55.50 & 61.17 & 68.74
& 55.94 & 58.47 & 60.96
& 37.93 & 43.68 & 52.87
\\

& \textbf{+ Ours}
& \textbf{59.98} & \textbf{67.51} & \textbf{71.43}
& \textbf{58.16} & \textbf{61.37} & \textbf{62.56}
& \textbf{40.23} & \textbf{51.72} & \textbf{55.17}
\\

& $\Delta$
& $+4.48$ & $+6.34$ & $+2.69$
& $+2.22$ & $+2.90$ & $+1.60$
& $+2.30$ & $+8.04$ & $+2.30$
\\

\bottomrule
\end{tabular}
}
\end{table}

\begin{table}[H]
\caption{
Complete SkillRet retrieval results (\%); per-backbone best results are in \textbf{bold}.
}
\label{tab:retrieval_skillret_full}
\centering

{
\small
\setlength{\tabcolsep}{0.35pt}
\renewcommand{\arraystretch}{0.88}
\setlength{\dashlinedash}{1.6pt}
\setlength{\dashlinegap}{1.2pt}
\setlength{\arrayrulewidth}{0.4pt}

\begin{tabular}{
    @{}
    >{\raggedright\arraybackslash}p{3.75cm}
    >{\centering\arraybackslash}p{1.10cm}
    *{9}{>{\centering\arraybackslash}p{\dimexpr(\textwidth-4.85cm-20\tabcolsep)/9\relax}}
    @{}
}
\toprule

\multirow{2}{*}{\textbf{Retriever}}
& \multirow{2}{*}{\textbf{Setting}}
& \multicolumn{3}{c}{\textbf{Recall}}
& \multicolumn{3}{c}{\textbf{NDCG}}
& \multicolumn{3}{c}{\textbf{Comp.}}
\\
\cmidrule(lr){3-5}
\cmidrule(lr){6-8}
\cmidrule(lr){9-11}
& & \textbf{@5} & \textbf{@10} & \textbf{@15}
& \textbf{@5} & \textbf{@10} & \textbf{@15}
& \textbf{@5} & \textbf{@10} & \textbf{@15}
\\

\midrule
\multicolumn{11}{c}{\textbf{Sparse Retrieval}}
\\
\cmidrule{1-11}

BM25
& Original
& 49.66 & 56.17 & 59.71
& 46.06 & 48.47 & 49.54
& 34.48 & 40.70 & 44.19
\\

\midrule
\multicolumn{11}{c}{\textbf{General Embedding Retrievers}}
\\
\cmidrule{1-11}

BGE-large-en-v1.5
& Original
& 55.96 & 61.40 & 64.31
& 53.75 & 55.81 & 56.71
& 39.30 & 44.63 & 47.49
\\

E5-large-v2
& Original
& 51.51 & 57.42 & 60.86
& 48.01 & 50.21 & 51.28
& 36.54 & 41.95 & 45.17
\\

Qwen3-Embedding-8B
& Original
& 60.64 & 67.10 & 70.39
& 57.52 & 59.98 & 61.01
& 43.33 & 49.99 & 53.53
\\

\multirow[c]{3}{*}{Qwen3-Embedding-0.6B}
& Original
& 59.14 & 64.91 & 68.20
& 56.21 & 58.40 & 59.42
& 41.83 & 47.21 & 50.47
\\

& \textbf{+ Ours}
& \textbf{66.93} & \textbf{71.73} & \textbf{74.48}
& \textbf{60.89} & \textbf{62.69} & \textbf{63.52}
& \textbf{51.59} & \textbf{56.95} & \textbf{60.16}
\\

& $\Delta$
& $+7.79$ & $+6.82$ & $+6.28$
& $+4.68$ & $+4.29$ & $+4.10$
& $+9.76$ & $+9.74$ & $+9.69$
\\

\midrule
\multicolumn{11}{c}{\textbf{Skill-oriented Retrievers}}
\\
\cmidrule{1-11}

\multirow[c]{3}{*}{SkillRouter-Embedding-0.6B}
& Original
& 71.01 & 76.00 & 78.71
& 68.88 & 70.82 & 71.67
& 53.47 & 59.32 & 62.60
\\

& \textbf{+ Ours}
& \textbf{75.54} & \textbf{81.00} & \textbf{83.49}
& \textbf{70.52} & \textbf{72.56} & \textbf{73.32}
& \textbf{61.50} & \textbf{68.44} & \textbf{71.72}
\\

& $\Delta$
& $+4.53$ & $+5.00$ & $+4.78$
& $+1.64$ & $+1.74$ & $+1.65$
& $+8.03$ & $+9.12$ & $+9.12$
\\

\noalign{\vskip 2pt}
\cdashline{1-11}[1.8pt/1.4pt]
\noalign{\vskip 2pt}

\multirow[c]{3}{*}{SkillRet-Embedding-0.6B}
& Original
& 79.14 & 85.33 & 88.01
& 75.58 & \textbf{78.01} & \textbf{78.84}
& 66.02 & 74.98 & 78.89
\\

& \textbf{+ Ours}
& \textbf{79.91} & \textbf{85.53} & \textbf{88.36}
& \textbf{75.70} & 77.87 & 78.75
& \textbf{67.54} & \textbf{75.71} & \textbf{79.77}
\\

& $\Delta$
& $+0.77$ & $+0.20$ & $+0.35$
& $+0.12$ & $-0.14$ & $-0.09$
& $+1.52$ & $+0.73$ & $+0.88$
\\

\noalign{\vskip 2pt}
\cdashline{1-11}[1.8pt/1.4pt]
\noalign{\vskip 2pt}

\multirow[c]{3}{*}{R3-Embedding}
& Original
& 81.82 & 87.64 & \textbf{90.20}
& 78.79 & \textbf{81.06} & \textbf{81.89}
& 68.78 & 77.43 & 81.33
\\

& \textbf{+ Ours}
& \textbf{82.76} & \textbf{87.81} & 90.19
& \textbf{78.85} & 80.83 & 81.57
& \textbf{70.70} & \textbf{78.31} & \textbf{81.95}
\\

& $\Delta$
& $+0.94$ & $+0.17$ & $-0.01$
& $+0.06$ & $-0.23$ & $-0.32$
& $+1.92$ & $+0.88$ & $+0.62$
\\

\bottomrule
\end{tabular}
}
\end{table}

\begin{table}[H]
\caption{
Complete SkillNet-Gym retrieval results (\%); per-backbone best results are in \textbf{bold}.
}
\label{tab:retrieval_skillnet_full}
\centering

{
\small
\setlength{\tabcolsep}{0.35pt}
\renewcommand{\arraystretch}{0.88}
\setlength{\dashlinedash}{1.6pt}
\setlength{\dashlinegap}{1.2pt}
\setlength{\arrayrulewidth}{0.4pt}

\begin{tabular}{
    @{}
    >{\raggedright\arraybackslash}p{3.75cm}
    >{\centering\arraybackslash}p{1.10cm}
    *{9}{>{\centering\arraybackslash}p{\dimexpr(\textwidth-4.85cm-20\tabcolsep)/9\relax}}
    @{}
}
\toprule

\multirow{2}{*}{\textbf{Retriever}}
& \multirow{2}{*}{\textbf{Setting}}
& \multicolumn{3}{c}{\textbf{Recall}}
& \multicolumn{3}{c}{\textbf{NDCG}}
& \multicolumn{3}{c}{\textbf{Comp.}}
\\
\cmidrule(lr){3-5}
\cmidrule(lr){6-8}
\cmidrule(lr){9-11}
& & \textbf{@5} & \textbf{@10} & \textbf{@15}
& \textbf{@5} & \textbf{@10} & \textbf{@15}
& \textbf{@5} & \textbf{@10} & \textbf{@15}
\\

\midrule
\multicolumn{11}{c}{\textbf{Sparse Retrieval}}
\\
\cmidrule{1-11}

BM25
& Original
& 48.90 & 60.64 & 64.26
& 45.17 & 50.70 & 52.02
& 22.64 & 42.77 & 46.54
\\

\midrule
\multicolumn{11}{c}{\textbf{General Embedding Retrievers}}
\\
\cmidrule{1-11}

BGE-large-en-v1.5
& Original
& 48.38 & 58.63 & 60.82
& 47.34 & 52.30 & 53.17
& 18.87 & 30.19 & 32.70
\\

E5-large-v2
& Original
& 43.11 & 51.01 & 55.20
& 41.61 & 45.44 & 47.07
& 14.47 & 24.53 & 29.56
\\

Qwen3-Embedding-8B
& Original
& 52.26 & 64.74 & 67.59
& 50.29 & 56.08 & 57.15
& 22.64 & 45.91 & 49.69
\\

\multirow[c]{3}{*}{Qwen3-Embedding-0.6B}
& Original
& 48.60 & 61.77 & 67.58
& 48.14 & 54.35 & 56.52
& 22.01 & 38.99 & 47.17
\\

& \textbf{+ Ours}
& \textbf{60.92} & \textbf{69.51} & \textbf{71.56}
& \textbf{58.82} & \textbf{62.96} & \textbf{63.71}
& \textbf{33.96} & \textbf{47.17} & \textbf{50.31}
\\

& $\Delta$
& $+12.32$ & $+7.74$ & $+3.98$
& $+10.68$ & $+8.61$ & $+7.19$
& $+11.95$ & $+8.18$ & $+3.14$
\\

\midrule
\multicolumn{11}{c}{\textbf{Skill-oriented Retrievers}}
\\
\cmidrule{1-11}

\multirow[c]{3}{*}{SkillRouter-Embedding-0.6B}
& Original
& 39.62 & 48.56 & 53.14
& 39.08 & 43.36 & 45.18
& 16.98 & 23.27 & 26.42
\\

& \textbf{+ Ours}
& \textbf{61.36} & \textbf{68.43} & \textbf{70.59}
& \textbf{59.40} & \textbf{62.74} & \textbf{63.56}
& \textbf{34.59} & \textbf{48.43} & \textbf{49.69}
\\

& $\Delta$
& $+21.74$ & $+19.87$ & $+17.45$
& $+20.32$ & $+19.38$ & $+18.38$
& $+17.61$ & $+25.16$ & $+23.27$
\\

\noalign{\vskip 2pt}
\cdashline{1-11}[1.8pt/1.4pt]
\noalign{\vskip 2pt}

\multirow[c]{3}{*}{SkillRet-Embedding-0.6B}
& Original
& 55.68 & 66.04 & 73.26
& 53.20 & 58.11 & 60.81
& 26.42 & 44.03 & 52.83
\\

& \textbf{+ Ours}
& \textbf{63.55} & \textbf{74.70} & \textbf{78.58}
& \textbf{60.16} & \textbf{65.33} & \textbf{66.72}
& \textbf{36.48} & \textbf{56.60} & \textbf{59.75}
\\

& $\Delta$
& $+7.87$ & $+8.66$ & $+5.32$
& $+6.96$ & $+7.22$ & $+5.91$
& $+10.06$ & $+12.57$ & $+6.92$
\\

\noalign{\vskip 2pt}
\cdashline{1-11}[1.8pt/1.4pt]
\noalign{\vskip 2pt}

\multirow[c]{3}{*}{R3-Embedding}
& Original
& 54.70 & 66.79 & 71.04
& 53.42 & 59.30 & 60.97
& 25.16 & 46.54 & 50.94
\\

& \textbf{+ Ours}
& \textbf{66.39} & \textbf{75.59} & \textbf{77.53}
& \textbf{62.47} & \textbf{66.82} & \textbf{67.54}
& \textbf{39.62} & \textbf{56.60} & \textbf{59.75}
\\

& $\Delta$
& $+11.69$ & $+8.80$ & $+6.49$
& $+9.05$ & $+7.52$ & $+6.57$
& $+14.46$ & $+10.06$ & $+8.81$
\\

\bottomrule
\end{tabular}
}
\end{table}
\subsection{Generalization to SRA-Bench}

To evaluate the generalizability of SkillDreamer across different skill retrieval benchmarks, we further conduct experiments on SRA-Bench. We apply SkillDreamer with Qwen3-Embedding-0.6B as the retrieval backbone and compare it with the original retriever. As shown in Table~\ref{tab:sra_results}, SkillDreamer improves Recall@5 from 57.90 to 66.08, NDCG@5 from 52.93 to 61.58, and Completeness@5 from 51.43 to 59.19, yielding gains of 8.18, 8.65, and 7.76 points, respectively. The consistent improvements across these three metrics show that SkillDreamer enhances retrieval coverage, ranking quality, and the completeness of the retrieved skill set. Similar gains are observed at $K=10$ and $K=15$, further demonstrating its effectiveness and generalizability on SRA-Bench.

\begin{table}[!htbp]
\caption{
Retrieval performance (\%) on SRA-Bench with Qwen3-Embedding-0.6B. Best results are highlighted in \textbf{bold}.
}
\label{tab:sra_results}
\centering

{
\small
\setlength{\tabcolsep}{1.5pt}
\renewcommand{\arraystretch}{1.0}
\setlength{\dashlinedash}{1.6pt}
\setlength{\dashlinegap}{1.2pt}
\setlength{\arrayrulewidth}{0.4pt}

\begin{tabular}{
    @{}
    >{\raggedright\arraybackslash}p{3.40cm}
    >{\centering\arraybackslash}p{1.15cm}
    *{9}{>{\centering\arraybackslash}p{\dimexpr(\textwidth-4.55cm-20\tabcolsep)/9\relax}}
    @{}
}
\toprule

\multirow{3}{*}{\textbf{Retriever}}
&
\multirow{3}{*}{\textbf{Setting}}
&
\multicolumn{3}{c}{\textbf{Recall}}
&
\multicolumn{3}{c}{\textbf{NDCG}}
&
\multicolumn{3}{c}{\textbf{Comp.}}
\\

\cmidrule(lr){3-5}
\cmidrule(lr){6-8}
\cmidrule(lr){9-11}

&
&
\textbf{@5}
&
\textbf{@10}
&
\textbf{@15}
&
\textbf{@5}
&
\textbf{@10}
&
\textbf{@15}
&
\textbf{@5}
&
\textbf{@10}
&
\textbf{@15}
\\

\midrule

\multirow[c]{3}{*}{Qwen3-Embedding-0.6B}

& Original
& 57.90 & 63.10 & 66.52
& 52.93 & 54.86 & 55.90
& 51.43 & 56.48 & 59.78
\\

&
\textbf{+ Ours}
&
\textbf{66.08}
&
\textbf{69.73}
&
\textbf{70.95}
&
\textbf{61.58}
&
\textbf{63.02}
&
\textbf{63.42}
&
\textbf{59.19}
&
\textbf{62.89}
&
\textbf{64.22}
\\

&
$\Delta$
&
$+8.18$
&
$+6.63$
&
$+4.43$
&
$+8.65$
&
$+8.16$
&
$+7.52$
&
$+7.76$
&
$+6.41$
&
$+4.44$
\\

\bottomrule

\end{tabular}
}

\end{table}
\FloatBarrier

\endgroup

\subsection{Generalization across Diverse Backbones}

To further verify the plug-and-play capability of SkillDreamer, we extend the evaluation to additional embedding backends, including BGE-large-en-v1.5~\citep{bge}, E5-large-v2~\citep{e5}, and Qwen3-Embedding-8B~\citep{qwen3embedding}. As shown in \cref{tab:additional_backbones_skillusage,tab:additional_backbones_skillret,tab:additional_backbones_skillnet}, SkillDreamer consistently improves Recall, NDCG, and Completeness across all backbones and datasets, demonstrating its effectiveness in enhancing skill coverage and ranking quality across diverse retrieval backbones. These results further validate that SkillDreamer can be seamlessly integrated with different embedding models without relying on a specific retrieval backbone.

\begin{table}[!htbp]
\caption{
Complete retrieval results (\%) with additional embedding backbones on SkillUsage; per-backbone best results are in \textbf{bold}.
}
\label{tab:additional_backbones_skillusage}
\centering
{
\small
\setlength{\tabcolsep}{0.35pt}
\renewcommand{\arraystretch}{0.88}
\setlength{\dashlinedash}{1.6pt}
\setlength{\dashlinegap}{1.2pt}
\setlength{\arrayrulewidth}{0.4pt}
\begin{tabular}{
    @{}
    >{\raggedright\arraybackslash}p{3.75cm}
    >{\centering\arraybackslash}p{1.10cm}
    *{9}{>{\centering\arraybackslash}p{\dimexpr(\textwidth-4.85cm-20\tabcolsep)/9\relax}}
    @{}
}
\toprule
\multirow{2}{*}{\textbf{Retriever}}
& \multirow{2}{*}{\textbf{Setting}}
& \multicolumn{3}{c}{\textbf{Recall}}
& \multicolumn{3}{c}{\textbf{NDCG}}
& \multicolumn{3}{c}{\textbf{Comp.}} \\
\cmidrule(lr){3-5}\cmidrule(lr){6-8}\cmidrule(lr){9-11}
& & \textbf{@5} & \textbf{@10} & \textbf{@15}
& \textbf{@5} & \textbf{@10} & \textbf{@15}
& \textbf{@5} & \textbf{@10} & \textbf{@15} \\
\midrule
\multirow[c]{3}{*}{BGE-large-en-v1.5}
& Original
& 47.56 & 53.88 & 56.84
& 48.24 & 50.92 & 51.98
& 32.18 & 35.63 & 39.08 \\
& \textbf{+ Ours}
& \textbf{58.83} & \textbf{63.38} & \textbf{65.37}
& \textbf{56.39} & \textbf{58.20} & \textbf{58.96}
& \textbf{40.23} & \textbf{45.98} & \textbf{48.28} \\
& $\Delta$
& $+11.27$ & $+9.50$ & $+8.53$
& $+8.15$ & $+7.28$ & $+6.98$
& $+8.05$ & $+10.35$ & $+9.20$ \\
\noalign{\vskip 2pt}
\cdashline{1-11}[1.8pt/1.4pt]
\noalign{\vskip 2pt}
\multirow[c]{3}{*}{E5-large-v2}
& Original
& 46.42 & 48.29 & 50.48
& 46.26 & 46.93 & 47.74
& 32.18 & 32.18 & 33.33 \\
& \textbf{+ Ours}
& \textbf{53.66} & \textbf{57.93} & \textbf{60.65}
& \textbf{53.43} & \textbf{55.14} & \textbf{56.15}
& \textbf{36.78} & \textbf{40.23} & \textbf{43.68} \\
& $\Delta$
& $+7.24$ & $+9.64$ & $+10.17$
& $+7.17$ & $+8.21$ & $+8.41$
& $+4.60$ & $+8.05$ & $+10.35$ \\
\noalign{\vskip 2pt}
\cdashline{1-11}[1.8pt/1.4pt]
\noalign{\vskip 2pt}
\multirow[c]{3}{*}{Qwen3-Embedding-8B}
& Original
& 53.39 & 60.95 & 69.72
& 54.77 & 57.91 & 60.78
& 32.18 & 41.38 & 50.57 \\
& \textbf{+ Ours}
& \textbf{58.42} & \textbf{69.11} & \textbf{73.45}
& \textbf{57.29} & \textbf{61.67} & \textbf{63.04}
& \textbf{37.93} & \textbf{50.57} & \textbf{55.17} \\
& $\Delta$
& $+5.03$ & $+8.16$ & $+3.73$
& $+2.52$ & $+3.76$ & $+2.26$
& $+5.75$ & $+9.19$ & $+4.60$ \\
\bottomrule
\end{tabular}
}
\end{table}

\begin{table}[!htbp]
\caption{
Complete retrieval results (\%) with additional embedding backbones on SkillRet; per-backbone best results are in \textbf{bold}.
}
\label{tab:additional_backbones_skillret}
\centering
{
\small
\setlength{\tabcolsep}{0.35pt}
\renewcommand{\arraystretch}{0.88}
\setlength{\dashlinedash}{1.6pt}
\setlength{\dashlinegap}{1.2pt}
\setlength{\arrayrulewidth}{0.4pt}
\begin{tabular}{
    @{}
    >{\raggedright\arraybackslash}p{3.75cm}
    >{\centering\arraybackslash}p{1.10cm}
    *{9}{>{\centering\arraybackslash}p{\dimexpr(\textwidth-4.85cm-20\tabcolsep)/9\relax}}
    @{}
}
\toprule
\multirow{2}{*}{\textbf{Retriever}}
& \multirow{2}{*}{\textbf{Setting}}
& \multicolumn{3}{c}{\textbf{Recall}}
& \multicolumn{3}{c}{\textbf{NDCG}}
& \multicolumn{3}{c}{\textbf{Comp.}} \\
\cmidrule(lr){3-5}\cmidrule(lr){6-8}\cmidrule(lr){9-11}
& & \textbf{@5} & \textbf{@10} & \textbf{@15}
& \textbf{@5} & \textbf{@10} & \textbf{@15}
& \textbf{@5} & \textbf{@10} & \textbf{@15} \\
\midrule
\multirow[c]{3}{*}{BGE-large-en-v1.5}
& Original
& 55.96 & 61.40 & 64.31
& 53.75 & 55.81 & 56.71
& 39.30 & 44.63 & 47.49 \\
& \textbf{+ Ours}
& \textbf{66.69} & \textbf{71.47} & \textbf{73.94}
& \textbf{61.42} & \textbf{63.23} & \textbf{63.98}
& \textbf{51.45} & \textbf{56.95} & \textbf{59.74} \\
& $\Delta$
& $+10.73$ & $+10.07$ & $+9.63$
& $+7.67$ & $+7.42$ & $+7.27$
& $+12.15$ & $+12.32$ & $+12.25$ \\
\noalign{\vskip 2pt}
\cdashline{1-11}[1.8pt/1.4pt]
\noalign{\vskip 2pt}
\multirow[c]{3}{*}{E5-large-v2}
& Original
& 51.51 & 57.42 & 60.86
& 48.01 & 50.21 & 51.28
& 36.54 & 41.95 & 45.17 \\
& \textbf{+ Ours}
& \textbf{63.85} & \textbf{68.99} & \textbf{71.47}
& \textbf{58.20} & \textbf{60.12} & \textbf{60.86}
& \textbf{48.79} & \textbf{54.29} & \textbf{57.17} \\
& $\Delta$
& $+12.34$ & $+11.57$ & $+10.61$
& $+10.19$ & $+9.91$ & $+9.58$
& $+12.25$ & $+12.34$ & $+12.00$ \\
\noalign{\vskip 2pt}
\cdashline{1-11}[1.8pt/1.4pt]
\noalign{\vskip 2pt}
\multirow[c]{3}{*}{Qwen3-Embedding-8B}
& Original
& 60.64 & 67.10 & 70.39
& 57.52 & 59.98 & 61.01
& 43.33 & 49.99 & 53.53 \\
& \textbf{+ Ours}
& \textbf{67.79} & \textbf{73.92} & \textbf{76.57}
& \textbf{62.06} & \textbf{64.35} & \textbf{65.16}
& \textbf{52.75} & \textbf{59.76} & \textbf{63.00} \\
& $\Delta$
& $+7.15$ & $+6.82$ & $+6.18$
& $+4.54$ & $+4.37$ & $+4.15$
& $+9.42$ & $+9.77$ & $+9.47$ \\
\bottomrule
\end{tabular}
}
\end{table}

\begin{table}[!htbp]
\caption{
Complete retrieval results (\%) with additional embedding backbones on SkillNet-Gym; per-backbone best results are in \textbf{bold}.
}
\label{tab:additional_backbones_skillnet}
\centering
{
\small
\setlength{\tabcolsep}{0.35pt}
\renewcommand{\arraystretch}{0.88}
\setlength{\dashlinedash}{1.6pt}
\setlength{\dashlinegap}{1.2pt}
\setlength{\arrayrulewidth}{0.4pt}
\begin{tabular}{
    @{}
    >{\raggedright\arraybackslash}p{3.75cm}
    >{\centering\arraybackslash}p{1.10cm}
    *{9}{>{\centering\arraybackslash}p{\dimexpr(\textwidth-4.85cm-20\tabcolsep)/9\relax}}
    @{}
}
\toprule
\multirow{2}{*}{\textbf{Retriever}}
& \multirow{2}{*}{\textbf{Setting}}
& \multicolumn{3}{c}{\textbf{Recall}}
& \multicolumn{3}{c}{\textbf{NDCG}}
& \multicolumn{3}{c}{\textbf{Comp.}} \\
\cmidrule(lr){3-5}\cmidrule(lr){6-8}\cmidrule(lr){9-11}
& & \textbf{@5} & \textbf{@10} & \textbf{@15}
& \textbf{@5} & \textbf{@10} & \textbf{@15}
& \textbf{@5} & \textbf{@10} & \textbf{@15} \\
\midrule
\multirow[c]{3}{*}{BGE-large-en-v1.5}
& Original
& 48.38 & 58.63 & 60.82
& 47.34 & 52.30 & 53.17
& 18.87 & 30.19 & 32.70 \\
& \textbf{+ Ours}
& \textbf{61.99} & \textbf{70.30} & \textbf{71.93}
& \textbf{58.41} & \textbf{62.25} & \textbf{62.95}
& \textbf{32.70} & \textbf{46.54} & \textbf{47.80} \\
& $\Delta$
& $+13.61$ & $+11.67$ & $+11.11$
& $+11.07$ & $+9.95$ & $+9.78$
& $+13.83$ & $+16.35$ & $+15.10$ \\
\noalign{\vskip 2pt}
\cdashline{1-11}[1.8pt/1.4pt]
\noalign{\vskip 2pt}
\multirow[c]{3}{*}{E5-large-v2}
& Original
& 43.11 & 51.01 & 55.20
& 41.61 & 45.44 & 47.07
& 14.47 & 24.53 & 29.56 \\
& \textbf{+ Ours}
& \textbf{59.71} & \textbf{68.71} & \textbf{71.10}
& \textbf{57.00} & \textbf{61.35} & \textbf{62.28}
& \textbf{34.59} & \textbf{47.17} & \textbf{50.31} \\
& $\Delta$
& $+16.60$ & $+17.70$ & $+15.90$
& $+15.39$ & $+15.91$ & $+15.21$
& $+20.12$ & $+22.64$ & $+20.75$ \\
\noalign{\vskip 2pt}
\cdashline{1-11}[1.8pt/1.4pt]
\noalign{\vskip 2pt}
\multirow[c]{3}{*}{Qwen3-Embedding-8B}
& Original
& 52.26 & 64.74 & 67.59
& 50.29 & 56.08 & 57.15
& 22.64 & 45.91 & 49.69 \\
& \textbf{+ Ours}
& \textbf{63.84} & \textbf{73.95} & \textbf{76.55}
& \textbf{60.03} & \textbf{64.73} & \textbf{65.66}
& \textbf{37.74} & \textbf{54.09} & \textbf{57.86} \\
& $\Delta$
& $+11.58$ & $+9.21$ & $+8.96$
& $+9.74$ & $+8.65$ & $+8.51$
& $+15.10$ & $+8.18$ & $+8.17$ \\
\bottomrule
\end{tabular}
}
\end{table}
\FloatBarrier

\subsection{Additional Ablation Results}

Tables~\ref{tab:ablation_skillusage_full} and~\ref{tab:ablation_skillret_full} extend the ablation study in the main paper by reporting Recall@$K$, NDCG@$K$, and Completeness@$K$ at all evaluated cutoffs on SkillUsage and SkillRet. Consistent with the main results, the two pseudo-skill-only variants do not outperform Direct Retrieval, showing that prospective execution information alone cannot replace the original task semantics. In contrast, PSG + HSR performs best under every metric and cutoff, improving R@5 over Direct Retrieval by 7.74 and 7.79 points, NDCG@5 by 4.79 and 4.68 points, and Comp.@5 by 6.90 and 9.76 points on SkillUsage and SkillRet, respectively. These results further confirm the benefit of combining task semantics with capability-conditioned prospective information in hybrid skill retrieval.

\begin{table}[!htbp]
\caption{Ablation results (\%) on SkillUsage. Best results are in \textbf{bold}.}
\label{tab:ablation_skillusage_full}
\centering
{
\small
\setlength{\tabcolsep}{0.35pt}
\renewcommand{\arraystretch}{0.88}
\setlength{\dashlinedash}{1.6pt}
\setlength{\dashlinegap}{1.2pt}
\setlength{\arrayrulewidth}{0.4pt}
\begin{tabular}{
    @{}
    >{\raggedright\arraybackslash}p{3.6cm}
    *{9}{>{\centering\arraybackslash}p{\dimexpr(\textwidth-3.6cm-18\tabcolsep)/9\relax}}
    @{}
}
\toprule
\multirow{2}{*}{\textbf{Setting}}
& \multicolumn{3}{c}{\textbf{Recall}}
& \multicolumn{3}{c}{\textbf{NDCG}}
& \multicolumn{3}{c}{\textbf{Comp.}} \\
\cmidrule(lr){2-4}\cmidrule(lr){5-7}\cmidrule(lr){8-10}
& \textbf{@5} & \textbf{@10} & \textbf{@15}
& \textbf{@5} & \textbf{@10} & \textbf{@15}
& \textbf{@5} & \textbf{@10} & \textbf{@15} \\
\midrule
Direct Retrieval
& 52.68 & 62.25 & 67.66
& 52.62 & 56.51 & 58.36
& 35.63 & 43.68 & 48.28 \\
\noalign{\vskip 2pt}
\cdashline{1-10}[1.8pt/1.4pt]
\noalign{\vskip 2pt}
Pseudo Skill w/o Cap.
& 48.51 & 58.69 & 63.75
& 49.29 & 53.74 & 55.37
& 31.03 & 37.93 & 43.68 \\
Pseudo-Skill Retrieval
& 48.85 & 59.00 & 65.48
& 48.87 & 53.40 & 55.53
& 27.59 & 40.23 & 48.28 \\
\textbf{PSG + HSR}
& \textbf{60.42} & \textbf{69.47} & \textbf{70.57}
& \textbf{57.41} & \textbf{60.83} & \textbf{61.33}
& \textbf{42.53} & \textbf{51.72} & \textbf{52.87} \\
\bottomrule
\end{tabular}
}
\end{table}

\begin{table}[!htbp]
\caption{Ablation results (\%) on SkillRet. Best results are in \textbf{bold}.}
\label{tab:ablation_skillret_full}
\centering
{
\small
\setlength{\tabcolsep}{0.35pt}
\renewcommand{\arraystretch}{0.88}
\setlength{\dashlinedash}{1.6pt}
\setlength{\dashlinegap}{1.2pt}
\setlength{\arrayrulewidth}{0.4pt}
\begin{tabular}{
    @{}
    >{\raggedright\arraybackslash}p{3.6cm}
    *{9}{>{\centering\arraybackslash}p{\dimexpr(\textwidth-3.6cm-18\tabcolsep)/9\relax}}
    @{}
}
\toprule
\multirow{2}{*}{\textbf{Setting}}
& \multicolumn{3}{c}{\textbf{Recall}}
& \multicolumn{3}{c}{\textbf{NDCG}}
& \multicolumn{3}{c}{\textbf{Comp.}} \\
\cmidrule(lr){2-4}\cmidrule(lr){5-7}\cmidrule(lr){8-10}
& \textbf{@5} & \textbf{@10} & \textbf{@15}
& \textbf{@5} & \textbf{@10} & \textbf{@15}
& \textbf{@5} & \textbf{@10} & \textbf{@15} \\
\midrule
Direct Retrieval
& 59.14 & 64.91 & 68.20
& 56.21 & 58.40 & 59.42
& 41.83 & 47.21 & 50.47 \\
\noalign{\vskip 2pt}
\cdashline{1-10}[1.8pt/1.4pt]
\noalign{\vskip 2pt}
Pseudo Skill w/o Cap.
& 49.56 & 55.15 & 57.91
& 46.54 & 48.59 & 49.44
& 33.20 & 37.62 & 40.00 \\
Pseudo-Skill Retrieval
& 55.56 & 62.60 & 66.23
& 48.91 & 51.51 & 52.61
& 40.48 & 47.53 & 51.63 \\
\textbf{PSG + HSR}
& \textbf{66.93} & \textbf{71.73} & \textbf{74.48}
& \textbf{60.89} & \textbf{62.69} & \textbf{63.52}
& \textbf{51.59} & \textbf{56.95} & \textbf{60.16} \\
\bottomrule
\end{tabular}
}
\end{table}
\FloatBarrier

\subsection{Additional Analysis on the Number of Skills}

\begin{figure}[!htbp]
    \centering
    \includegraphics[width=0.46\linewidth]{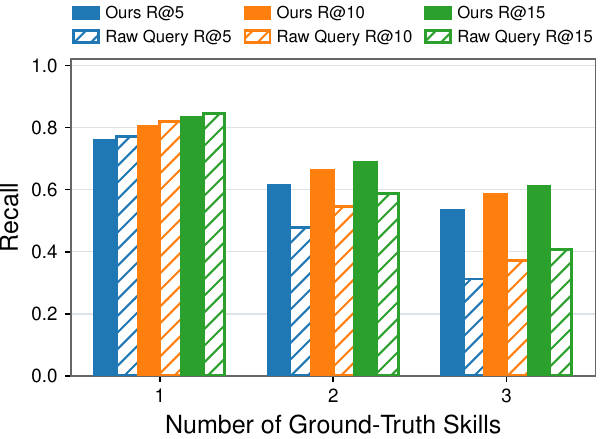}
    \caption{Recall@K on SkillRet across tasks with different numbers of ground-truth skills.}
    \label{fig:skillret_recall_by_gt_count}
\end{figure}

To further investigate the effect of the number of required skills, we extend the analysis to SkillRet. As shown in
Fig.~\ref{fig:skillret_recall_by_gt_count}, SkillDreamer maintains comparable retrieval performance with raw query retrieval for single-skill queries. As the number of ground-truth skills increases, SkillDreamer increasingly outperforms raw query retrieval across R@5, R@10, and R@15, achieving larger performance gains on more complex multi-skill queries. These results further demonstrate the effectiveness of SkillDreamer in identifying diverse skills required for complex tasks.

\FloatBarrier



\clearpage
\addcontentsline{toc}{section}{References}
\putbib
\end{bibunit}

\end{document}